\documentclass[11pt,aps,tightenlines,nofootinbib]{revtex4}

\usepackage{amsmath}
\usepackage{amssymb}
\usepackage{bbm}
\usepackage{mathtools}
\usepackage[normalem]{ulem}
\usepackage{dsfont}
\usepackage{mathrsfs}
\usepackage{cancel}
\usepackage{bbold}

\usepackage{braket}

\usepackage[utf8]{inputenc}
\usepackage[T1]{fontenc}

\def\eq{{eq}}

\def\la{\langle}
\def\ra{\rangle}

\def\ii{\mathrm{i}}
\def\cc{\text{c.c.}}
\def\idots{{i_1 i_2\dots i_N}}
\def\jdots{{j_1 j_2\dots j_N}}

\def\B#1{\!\left(#1\right)}
\def\BB#1{\!\left[#1\right]}

\def\be{\begin{equation}}
\def\ee{\end{equation}}

\def\bee{\begin{equation*}}
\def\eee{\end{equation*}}

\def\ba{\begin{equation}\begin{aligned}}
\def\ea{\end{aligned}\end{equation}}

\def\pref{0.65}

\def\fit{\text{fit}}
\def\GS{{GS}}
\def\EX{{EX}}

\def\for{\ \text{for} \ }

\def\cross{\text{cross}}

\def\two#1#2{\langle #1 \, {#2}\rangle}

\makeatletter
\newcommand{\pushright}[1]{\ifmeasuring@#1\else\omit\hfill$\displaystyle#1$\fi\ignorespaces}
\newcommand{\pushleft}[1]{\ifmeasuring@#1\else\omit$\displaystyle#1$\hfill\fi\ignorespaces}
\makeatother

\usepackage[ddmmyyyy]{datetime}
\makeatletter
\def\Dated@name{}
\makeatother

\begin{document}

\title{Dynamics of  longitudinal magnetization in   transverse-field quantum
Ising model: from  symmetry-breaking gap to   Kibble-Zurek mechanism}
\author{Micha{\l} Bia{\l}o{\'{n}}czyk and Bogdan Damski}
\affiliation{Jagiellonian University, Marian Smoluchowski Institute of Physics, {\L}ojasiewicza 11, 30-348 Krak\'ow, Poland}
\begin{abstract}
We show that   the symmetry-breaking gap of the quantum Ising model in the transverse field can be extracted
from  free evolution  of the longitudinal magnetization taking
place after a gradual  quench of the magnetic field. 
We perform  for this purpose  numerical simulations  of   the  Ising chains
with either periodic or open boundaries.
We  also study  the condition for adiabaticity of evolution of the longitudinal magnetization
 finding excellent agreement between 
 our simulations and  the prediction based on the Kibble-Zurek theory of non-equilibrium
phase transitions. Our results should be relevant for ongoing cold atom and
ion experiments targeting either  equilibrium or dynamical aspects of  
quantum phase transitions. They could be also useful for benchmarking D-Wave
machines.
\end{abstract}
\date{\today}
\maketitle

\section{Introduction}
\label{Introduction_sec}

Quantum phase transitions take place when tiny changes of the external
parameter, such as the magnetic field in spin models or the lattice height 
in cold atom setups, can induce radical changes in  ground state properties of the
system \cite{Sachdev,ContinentinoBook,SachdevToday,LewensteinAdv}. This can happen when the system is near the 
critical point separating its   phases.

One of the typical features associated with quantum phase transitions is the
symmetry-breaking phenomenon, where one of the phases of the
thermodynamically-large system has degenerate 
ground states, which do not respect the symmetry of
the Hamiltonian. 
Such a degeneracy is typically lifted  in
finite systems, where a small energy gap between  lowest-energy eigenstates of
the Hamiltonian is present. Rapid disappearance of this gap with increase of the system size signals the
onset of the symmetry-breaking phenomenon that is fundamentally important for
understanding of both equilibrium and non-equilibrium  phase
transitions.
It is thus  very  interesting  to address the question how one can
access the symmetry-breaking gap in real systems to observe emergence of such a compelling  phenomenon.

We make a step forward in this direction by providing an explicit two-stage scheme for
studies of the symmetry-breaking gap in 
the paradigmatic system undergoing a quantum  phase transition: the quantum Ising model in the
transverse magnetic field \cite{Sachdev}.
In the first stage, the system is gradually driven to the desired value of
the transverse magnetic field 
(see \cite{JacekPRL2005,DornerPRL2005,PolkovnikovPRB2005,RalfPRA2007,PolkovnikovPRL2008,SenPRB2009,SantoroPRB2009,JacekPRA2007,SenPRA2009,ArnabPRB2010,KolodrubetzPRL2012,FrancuzPRB2016,SantoroJstat2015,PuskarovSciP2016,ApollaroSciRep2017,MichalOne,AdolfoPRL2018,MarekArxiv} for   studies of  different aspects of dynamics of the
quantum Ising model under similar driving).  In the second stage, the system undergoes free
 evolution  and oscillations of its  longitudinal magnetization are recorded.
They  encode the symmetry-breaking gap.

While our studies target a specific model, we expect that they 
can be generalized to other, not necessarily 
exactly solvable systems, which can be experimentally approached in cold atom and
ion simulators of various  condensed matter models
\cite{LewensteinAdv,BlochSci2017}.
In fact, while pursuing  this work, we have come across a recent  paper
 discussing experimental studies of the symmetry-breaking gap in a
cold atom cloud \cite{LMGexp}. We will first present 
 our results and then compare the  two approaches. 

The outline of this paper is the following. We explain the idea behind our
work in Sec. \ref{Idea_sec}. 
We  illustrate in
Sec. \ref{Two_sec} the
concepts related to the symmetry-breaking gap and the 
equilibrium longitudinal magnetization in the simplest version of the Ising model composed of just two
spins.  We then discuss in Sec. \ref{Periodic_sec} dynamical extraction of those quantities from after-quench 
free evolution of the longitudinal magnetization in the periodic
Ising chain composed of several spins. Next, we extend these studies in Sec.
\ref{Open_sec} to systems composed of up to several hundreds of spins by
considering Ising chains with open boundaries, where  computations of the longitudinal
magnetization can be more efficiently done. The overall
discussion of our results is presented in Sec. \ref{Discussion_sec}. 
Finally, technicalities related to the  studies of the  Ising
chains with periodic (open) boundaries  are presented in Appendix \ref{Mx_app} (Appendix \ref{OBC_app}).

\section{Idea}
\label{Idea_sec}

To explain the idea behind our proposal, we introduce the 
quantum Ising model in the transverse magnetic field $g$, whose Hamiltonian 
in the periodic chain can be written as \cite{Lieb1961}
\be
\begin{aligned}
& H(g) = -\sum_{i=1}^{N}\B{\sigma^x_i\sigma^x_{i+1} + g \sigma^z_i},\\
&\sigma^x_{N+1}=\sigma^x_{1},
\end{aligned}
\label{PBC}%
\ee
where $\sigma_j^{x,y,z}$ is the Pauli matrix acting on the $j$-th spin,  $N$
is the number of spins, and   $g\ge0$ is assumed in this work. 
Such a model has two phases: the ferromagnetic phase for  $0\le g<1$ and 
the paramagnetic phase for $g>1$.
The breaking of the $\mathbb{Z}_2$ symmetry of this model, associated with
the $\sigma^x_j\to-\sigma^x_j$ symmetry of the Hamiltonian, 
can be easily explained  at $g=0$, when the two ground states have all
spins aligned in the $\pm x$ direction 
\be
|\rightarrow\rightarrow\rightarrow\cdots\ra, \ |\leftarrow\leftarrow\leftarrow\cdots\ra.
\label{twodeg}
\ee
Then an arbitrarily  small perturbation  along the $x$ direction, say
$-h\sum_i\sigma^x_i$ with $h\to0^+$, makes 
$|\rightarrow\rightarrow\rightarrow\cdots\ra$ the ground state of the system breaking the spin-flip symmetry
supported by the Hamiltonian. Importantly, degeneracy of the ground state, in thermodynamically-large 
systems, persists in
the whole ferromagnetic phase.

The idea that we have is that one can initially set $g=0$ and  prepare the system in one of the states 
given in (\ref{twodeg}), say  
\be
|\psi\ra= |\rightarrow\rightarrow\rightarrow\cdots\ra.
\label{initial}
\ee
We propose then to evolve the system according to the following  protocol 
\be
g(t)=
\left\{
\begin{array}{ll}
g_f+t/\tau_Q &   \text{for} \  -g_f\tau_Q  \le t  \le  0\\
g_f   &    \text{for} \  t>0
\end{array}
\right.
\label{godt}
\ee
with $\tau_Q$ controlling the rate of  driving   the system to the
magnetic field $g_f$, where  free evolution begins.

\begin{figure}[t]
\includegraphics[width=\pref\textwidth,clip=true]{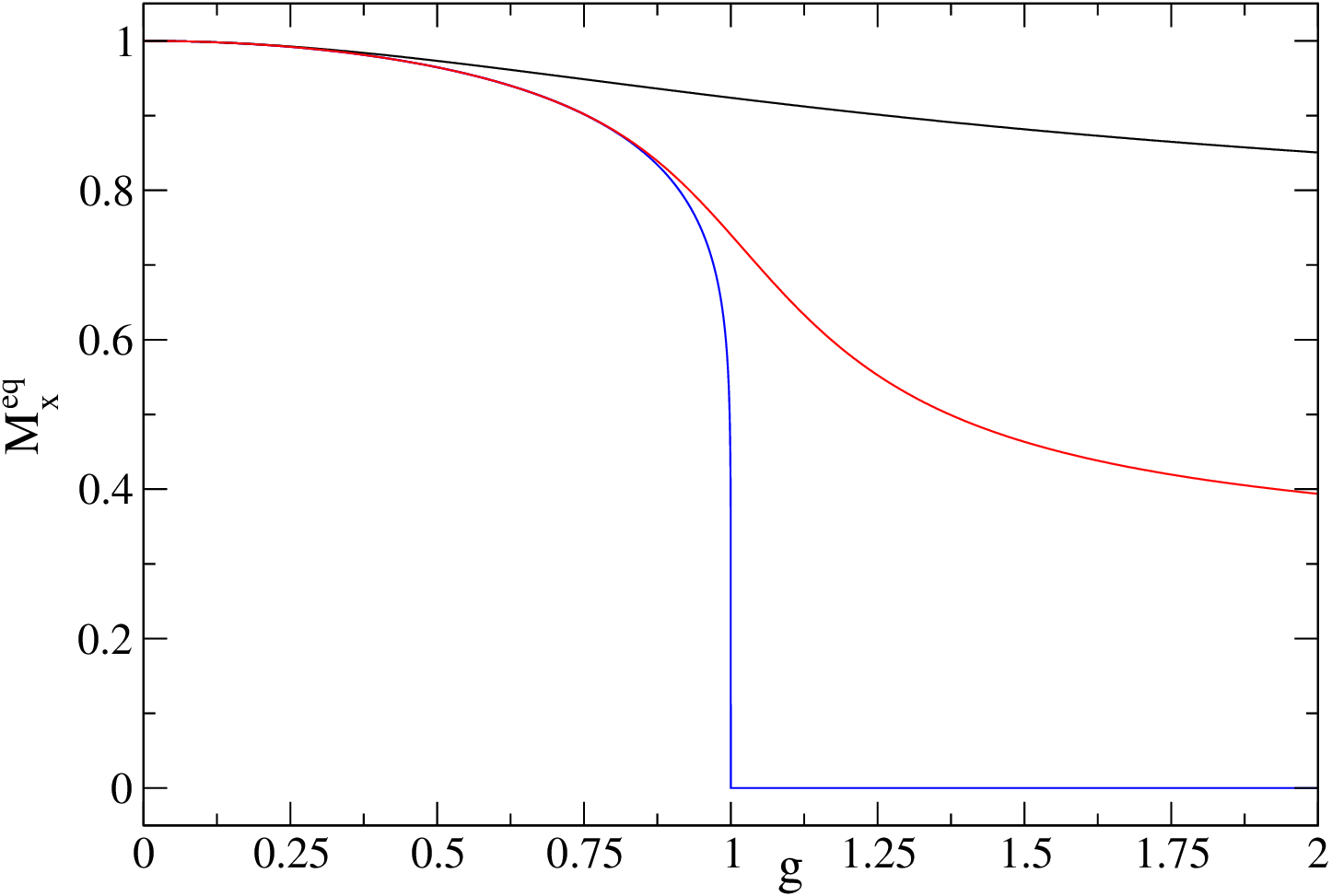}
\caption{
The  equilibrium longitudinal magnetization $M_x^\eq$ of  periodic  Ising chains. 
The curves, top to bottom, correspond to system sizes $2$ (black), $12$
(red),  and $\infty$ (blue).
The first of them comes from  (\ref{Mx2}), the second  one 
has been numerically obtained through exact diagonalization,  the last one 
is given by   (\ref{Onsager}).
}
\label{Mxeq_fig}
\end{figure}

The natural observable for  studies of the symmetry-breaking gap is 
the longitudinal magnetization
\be
M_x=\la\sigma^x_i\ra,
\label{kkqq}
\ee
where the choice of the site index $i$ is arbitrary in a periodic chain.
Such an observable is sensitive to the way in which the symmetry is broken.
 Indeed, it is well-known that 
\be
-M_x^\eq \le M_x\le M_x^\eq,
\ee
where in the thermodynamically-large system  \cite{Pfeuty}
\be
\ M^\eq_x=\left\{
\begin{array}{ll}
(1-g^2)^{1/8} &  \text{for} \  0\le g\le1\\
0     &  \text{for} \  g>1
\end{array}
\right..
\label{Onsager}
\ee
$M_x^\eq$ will be called the equilibrium longitudinal magnetization. It was
obtained in the $N\to\infty$ limit from  the asymptotic form of the two-point correlation functions.
We will study it in finite systems, where a different approach has to be employed (see below).
Typical results that we obtain are illustrated in Fig. \ref{Mxeq_fig}, where we  see that 
$M_x^\eq$ is analytic when $N<\infty$. In such systems,
the  drop of the longitudinal magnetization is not as
steep as in (\ref{Onsager}) on the ferromagnetic side.
On the paramagnetic side,
non-zero value of $M_x^\eq$ appears--see   (\ref{Mxinfty}) and discussion around
it.

The question now is how the symmetry-breaking gap can be extracted from
dynamics of the longitudinal magnetization. To explain that, we first note 
 that  the expectation value of the $\sigma^x_j$ operator vanishes 
 in 
 all eigenstates of  Hamiltonian 
(\ref{PBC}) for $g>0$. This  can be proved with Wick's theorem \cite{Wick} in the fermionic representation 
of the Ising model, which we briefly introduce  in Appendix \ref{Mx_app}.
Non-zero value of the longitudinal magnetization  
can be  obtained by evaluation of  (\ref{kkqq}) in a proper  superposition of  two  eigenstates 
of the Hamiltonian.
In fact,  deceptively simple-looking initial state (\ref{initial}) is a macroscopic
superposition of two  eigenstates belonging to  different sectors of the spectrum of
Hamiltonian (\ref{PBC}).

To understand this statement, we note that   Hamiltonian
(\ref{PBC}) commutes with the parity operator,
\be
P=\prod_{i=1}^N \sigma^z_i,  \   [H,P]=0.
\label{HP}
\ee
This leads to splitting of the Hilbert space into two subspaces, where the eigenstates have either
$+1$ or $-1$ parity \cite{BDJPA2014}. The  symmetry-breaking gap $\delta$ is the  difference
between the energies of the lowest-energy eigenstates    in the negative and positive parity subspaces
in the ferromagnetic phase, where the symmetry-breaking phenomenon takes place.
Such a quantity, however, will be of key importance in 
our studies  for any value of
the magnetic field $g$. We will thus call it the symmetry-breaking gap
even when it will be computed at the critical point or in the paramagnetic
phase, which  will simplify our discussion.

\begin{figure}[t]
\includegraphics[width=\pref\textwidth,clip=true]{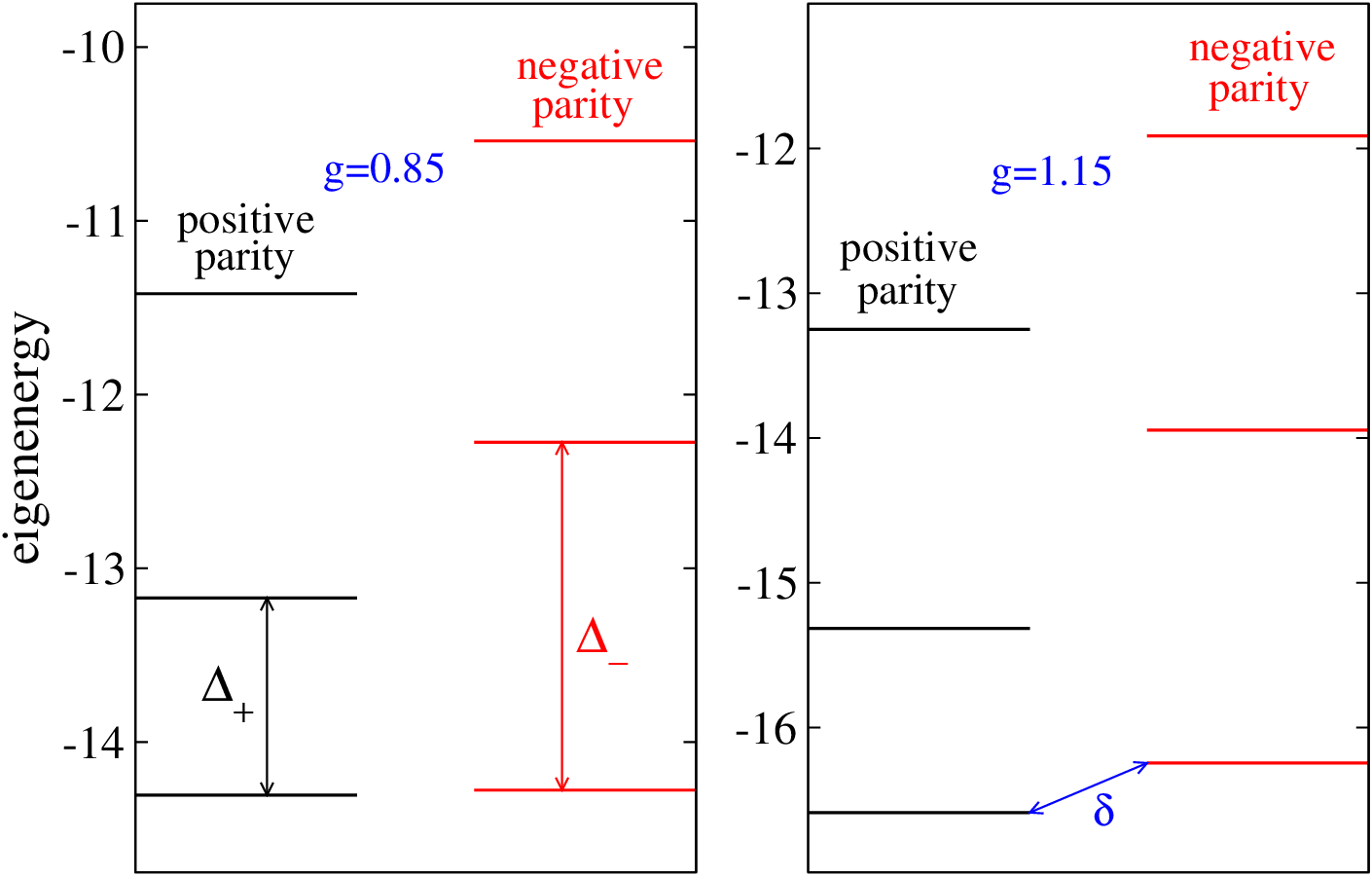}
\caption{Lowest-energy levels of Hamiltonian (\ref{PBC}) for $N=12$, which are populated during 
time evolution starting from state (\ref{initial}) and driven  by quench protocol  (\ref{godt}). 
Left panel: typical results in the
ferromagnetic phase. The positive and negative parity ground states are 
nearly degenerate despite the relatively small system size. 
Right panel: typical results in the paramagnetic phase. 
The dynamical gaps in the positive and negative parity subspaces,
$\Delta_+$ and $\Delta_-$, respectively, and the symmetry-breaking gap
$\delta$ are marked to illustrate  key quantities involved in our
studies. Their values for the parameters used in this figure can be found in
Table \ref{tab_eq}. 
}
\label{levels_fig}
\end{figure}

These features of the spectrum are illustrated in Fig. \ref{levels_fig}.
A closed-form expression for $\delta$  was given in  \cite{BDJPA2014}
\begin{align}
&\delta  =  
g^N \int_0^1 dt\, \frac{4 N}{\pi} \frac{t^{N-3/2} \sqrt{(1-t)(1-g^2 t)} }{1-(gt)^{2N} } \  {\rm for} \  0\le g<1,
\label{gap} \\
&\delta = 2g-2 +   g^{-N} \int_0^1 dt\,\frac{4 N}{\pi} 
\frac{t^{N-3/2} \sqrt{(1-t)(g^2-t)} }{1- t^{2N}/g^{2N}} \ {\rm for}  \ g>1,
\label{gappara}\\
&\delta = 2\tan\B{\frac{\pi}{4N}}  \ {\rm at} \  g=1.
\label{gapcritical}
\end{align}
These expressions can be used for showing that
$\delta$ vanishes exponentially (algebraically) with  the system size deeply in the ferromagnetic
phase (near the critical point) \cite{BDJPA2014}. Deeply in the paramagnetic phase, $\delta$ is
well-approximated by an expression that is obtained from (\ref{gappara}) after 
neglecting the term containing  the integral. 

Besides the symmetry-breaking gap, there are  also  dynamical gaps
$\Delta_+$ and  $\Delta_-$. They are  defined as energy gaps 
in  subspaces of positive and negative parity 
between the ground state and the first
excited state that can be
populated during evolution starting from (\ref{initial})--see Fig.
\ref{levels_fig}.
Their importance comes from the fact that commutation relation (\ref{HP}) 
prohibits dynamical transitions between the two parity  subspaces. As a
result, system's excitation due to  driving (\ref{godt}) 
depends on $\Delta_\pm$, the quench time $\tau_Q$, 
and the initial state  (\ref{initial}) for time evolution. It  
has nothing to do with the symmetry-breaking gap $\delta$ as long as there is
no symmetry-breaking perturbation in the system, which is the case in our
studies.

If we now note that $\sigma^z|\rightarrow\ra=|\leftarrow\ra$, which implies 
$\sigma^z|\leftarrow\ra=|\rightarrow\ra$, we see that the ground states 
in the positive and negative parity subspaces at $g=0$ can be written as 
\be
|\GS_\pm(g=0)\ra=\frac{|\rightarrow\rightarrow\rightarrow\cdots\ra \pm |\leftarrow\leftarrow\leftarrow\cdots\ra
}{\sqrt{2}},
\ee
which in turn allows us to cast initial state (\ref{initial})  to the following form
\be
|\psi\ra=\frac{|\GS_+(g=0)\ra+|\GS_-(g=0)\ra}{\sqrt{2}}.
\ee

Next, we use (\ref{HP}) and employ time-dependent Schr\"odinger equation to arrive at    
\be
\begin{aligned}
&|\psi(t)\ra=|\psi_+(t)\ra + |\psi_-(t)\ra,\\
&P|\psi_\pm(t)\ra=\pm|\psi_\pm(t)\ra.
\end{aligned}
\label{psit}
\ee
Finally, knowing that $\sigma^x_i$ has vanishing matrix elements in  the positive and
negative parity subspaces, which can be easily proven with the help of Wick's theorem, 
we get 
\be
M_x(t)=\la\psi(t)|\sigma^x_i|\psi(t)\ra=
\la\psi_-(t)|\sigma^x_i|\psi_+(t)\ra + \text{c.c.}
\ee

\begin{table}[t]
\begin{tabular}{ l | c | c | c | c}
  &  $\delta$ & $\Delta_+$ & $\Delta_-$ & $M^\eq_x$  \\
  \hline
  $g=0.85$ &  0.028747       &    1.13    & 2.00  & 0.8545     	 \\
  \hline  
  $g=1$ &  0.13109     &  1.04    	&  2.07   &  0.7407  \\
  \hline
  $g=1.15$ & 0.34167   &  1.27   	 &  2.30  & 0.6146
\end{tabular}
\caption{The symmetry-breaking gap $\delta$, the dynamical gaps
$\Delta_\pm$, and the
equilibrium longitudinal magnetization $M^\eq_x$  in the periodic Ising chain  composed of $N=12$
spins. The first column is  obtained from (\ref{gap})--(\ref{gapcritical}),
 the next two columns come from   (\ref{DD}), the fourth column is
 numerically obtained through exact diagonalization.   }
\label{tab_eq}
\end{table}

This  expression can be very  complicated for fast quenches, i.e.,
for small $\tau_Q$ in (\ref{godt}).
However, if we  assume that the driving  is slow enough to be nearly 
adiabatic, then the following expression will properly approximate the exact
result
\be
M_x(t)=\cos\B{\int_{-g_f\tau_Q}^t dt\, \delta[g(t)]}\la\GS_-[g(t)]|\sigma^x_i|\GS_+[g(t)]\ra.
\label{Mxeqq}
\ee
Such an expression is  still  non-trivial because we are unaware of a closed-form expression for
the above matrix element in an arbitrarily-sized system.
If we now consider $t>0$, i.e., the free evolution stage in our problem, we
will get from (\ref{Mxeqq}) that 
\be
M_x(t)=\cos\B{\delta(g_f) t+\text{const}} M^\eq_x(g_f),
\label{Mxeqq_bis}
\ee
which can be used for extracting the symmetry-breaking gap and the equilibrium
longitudinal magnetization out of either  numerical or experimental data.

The question now is what is the condition for adiabaticity in our system so
that approximation (\ref{Mxeqq}) can be used.
Clearly, the system is most prone to being excited near the critical point,
where the dynamical gap $\Delta_\pm$ 
is smallest,   and the quantum version \cite{BDPRL2005,DornerPRL2005,JacekPRL2005} of the
Kibble-Zurek (KZ) theory of non-equilibrium phase transitions
\cite{KibbleRev,ZurekRev,delCampoReva,delCampoRevb} can be used. 
A simple criteria then exists and is based on comparision between the system
size and the non-equilibrium  length-scale $\hat\xi$ characterizing excitations resulting from the
quench \cite{JacekAdv2010}. The latter, according to  the KZ theory, scales with the quench time as
\be
\hat\xi\sim \tau_Q^{\nu/(1+z\nu)},
\ee
where $z$ and $\nu$ are the critical exponents.
These exponents  describe disappearance
of the dynamical gap and the correlation length, which are 
assumed to be proportional to $|g-g_c|^{z\nu}$ and  $|g-g_c|^{-\nu}$,
respectively, near the critical point $g_c$.
Comparing the two length scales, one gets the following  condition for a crossover between
adiabatic and non-adiabatic evolutions approaching or crossing the critical
point
\be
\tau_Q\sim N^{(1+z\nu)/\nu}\sim N^2,
\label{kzscaling}
\ee
where we substituted   $z=\nu=1$ for the quantum Ising model in the transverse
field. In other words, for quench times of the order of $N^2$ or larger we
expect the evolution to be nearly adiabatic.

\section{Two spins}
\label{Two_sec}
We illustrate here the quantities  introduced in the previous section by showing 
how the symmetry-breaking gap and the longitudinal magnetization can be calculated  in the
simplest version of the Ising chain, the one with just two spins. 

To start, we need eigenenergies and normalized  eigenstates of (\ref{PBC}),
which for $N=2$ read 
\begin{align}
&{\cal E}_{\GS_+}=-2\sqrt{1+g^2}, \
|\GS_+\ra=A_-|\uparrow\uparrow\ra+B_-|\downarrow\downarrow\ra, \\
&{\cal E}_{\GS_-}=-2, \ |\GS_-\ra=\frac{|\uparrow\downarrow\ra+|\downarrow\uparrow\ra}{\sqrt{2}}, \\
&{\cal E}_{\EX_-}=2,  \ |\EX_-\ra=\frac{|\uparrow\downarrow\ra-|\downarrow\uparrow\ra}{\sqrt{2}},\\
&{\cal E}_{\EX_+}=2\sqrt{1+g^2}, \ |\EX_+\ra=A_+|\uparrow\uparrow\ra+B_+|\downarrow\downarrow\ra,
\end{align}
where  $\sigma^z|\uparrow\ra=|\uparrow\ra$,
$\sigma^z|\downarrow\ra=-|\downarrow\ra$,
\be
A_\pm=\frac{1}{\sqrt{2}\sqrt{1+g^2\pm g\sqrt{1+g^2}}}, \ 
B_\pm=-\frac{g\pm\sqrt{1+g^2}}{\sqrt{2}\sqrt{1+g^2\pm g\sqrt{1+g^2}}},
\ee
and   $\EX_\pm$ refers to the  excited states in the
corresponding subspaces. 

Using these results, one immediately gets
\be
\delta={\cal E}_{\GS_-}-{\cal E}_{\GS_+}=2\B{\sqrt{1+g^2}-1},
\ee
which agrees with (\ref{gap})--(\ref{gapcritical}) for $N=2$.

It is then a standard exercise to show that the extremal values of 
the longitudinal magnetization in the state, which is an arbitrary
superposition of $|\GS_+\ra$ and $|\GS_-\ra$, are 
\be
\begin{aligned}
M_x&=\pm\la\GS_-|\sigma^x_i|\GS_+\ra=\pm M_x^\eq\\
&=\pm\frac{1}{2}\B{
\sqrt{1+\frac{g}{\sqrt{1+g^2}}}+
\sqrt{1-\frac{g}{\sqrt{1+g^2}}}
}.
\end{aligned}
\label{Mx2}
\ee

This formula  is depicted in Fig. \ref{Mxeq_fig}, where one easily
notices its departures from thermodynamic limit expression (\ref{Onsager}).
These differences do not vanish in the whole paramagnetic phase.
In fact, for a periodic system composed of $N$ spins, it is easy to show that the longitudinal magnetization 
 does not vanish  in the limit of $g\to\infty$ 
\be
M^\eq_x(g\to\infty)=\frac{1}{\sqrt{N}}.
\label{Mxinfty}
\ee
Such a simple result follows from elementary  observation that in the limit of $g\to\infty$ the
positive parity ground state  approaches 
$|\uparrow\uparrow\uparrow\uparrow\cdots\ra$, while the negative parity one
approaches 
\be
\frac{
|\downarrow\uparrow\uparrow\uparrow\cdots\ra+
|\uparrow\downarrow\uparrow\uparrow\cdots\ra+
|\uparrow\uparrow\downarrow\uparrow\cdots\ra+\cdots
}{\sqrt{N}}.
\ee

\section{Dynamics of  periodic chains}
\label{Periodic_sec}
We will discuss here dynamics of the longitudinal magnetization in  a periodic Ising  chain. 
There are different ways how one can approach this problem.

The most direct one is to work  
in the Fock space spanned by all combinations of up/down states of all spins  such as
$|\uparrow\downarrow\downarrow\uparrow\cdots\ra$. Such a  space can be
then easily cut into positive and negative parity subspaces, where 
time evolutions  can be independently performed.
The advantage of such an approach is that it allows for easy computation of
the longitudinal magnetization. The disadvantage is rather obvious: one has to
keep track of $2\times2^{N-1}$ amplitudes, which highly limits the available
system sizes.

Another approach relies on mapping the system onto non-interacting fermions
via the Jordan-Wigner transformation (\ref{JW}). This reduces the problem to
finding time evolution of $N-1$ two-level uncoupled systems (\ref{Bogt}), which can be
straightforwardly   numerically achieved. This is the most
efficient way of getting wave-function (\ref{psit}). The problem with this approach, however, 
lies in the complexity of the computation of the longitudinal magnetization,
which we  explain in Appendix \ref{Mx_app}.

\begin{figure}[t]
\includegraphics[width=\pref\textwidth,clip=true]{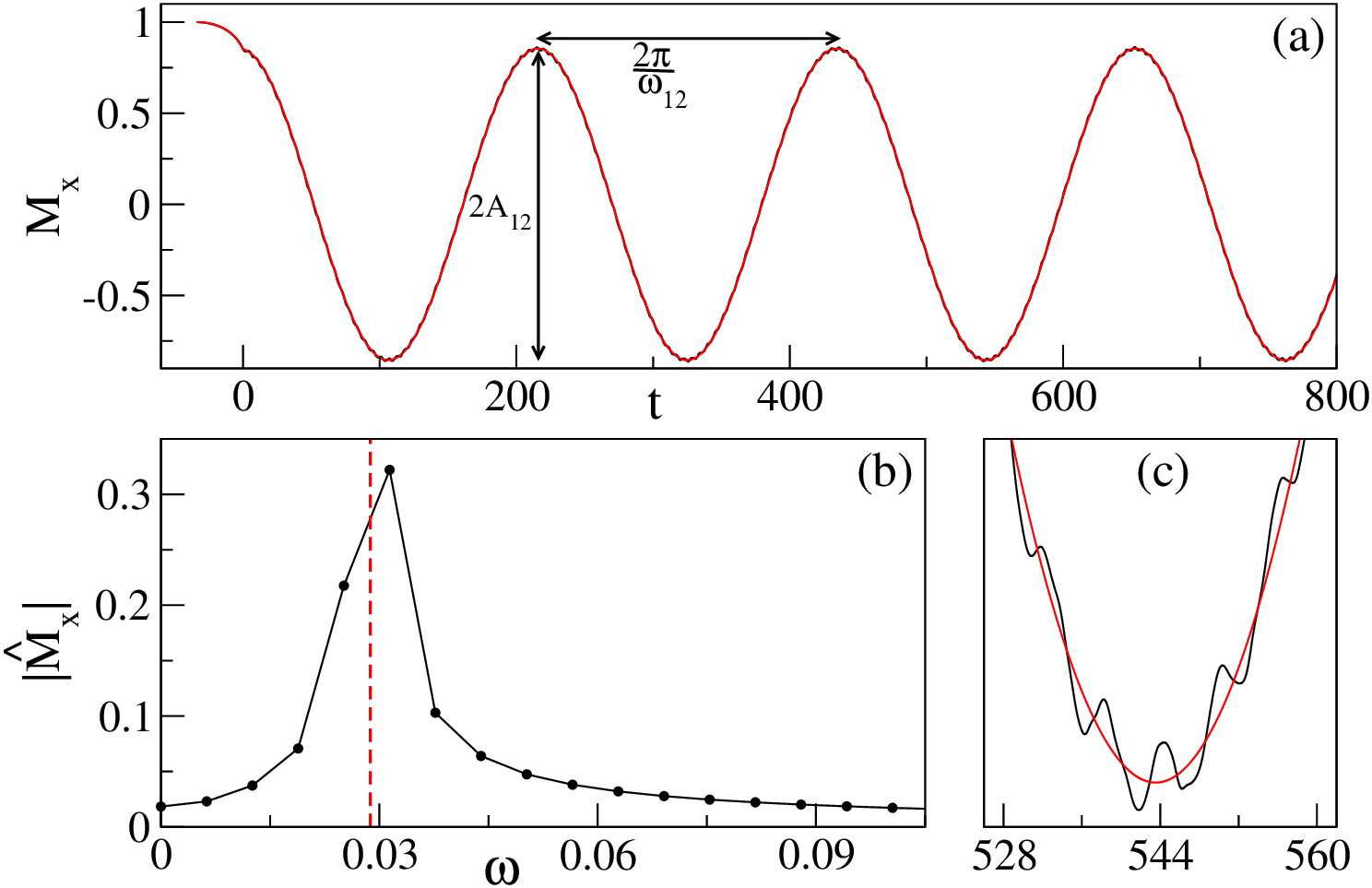}
\caption{Dynamics of the longitudinal magnetization in a periodic system for a nearly adiabatic 
quench that stops in the ferromagnetic phase at $g_f=0.85$. 
The driving ends   at the time $t=0$ and then free
evolution begins.
Panel (a): 
the black line shows numerics,
while the red one shows adiabatic approximation (\ref{Mxeqq}). 
Panel (b): modulus of the 
discrete Fourier transform (\ref{Fourier}) of  free evolution of the longitudinal magnetization around the
symmetry-breaking  gap $\delta$ 
marked by the vertical red dashed line.  Data points are joined by line segments to guide the eye. Panel (c): 
magnification of the area around the third minimum from panel (a).
The system size is $N=12$, the quench time is $\tau_Q=40$, and the time span of
free oscillations used for computing (\ref{Fourier})  is $L=1000$.
}
\label{per_0_85_fig}
\end{figure}

\begin{figure}[t]
\includegraphics[width=\pref\textwidth,clip=true]{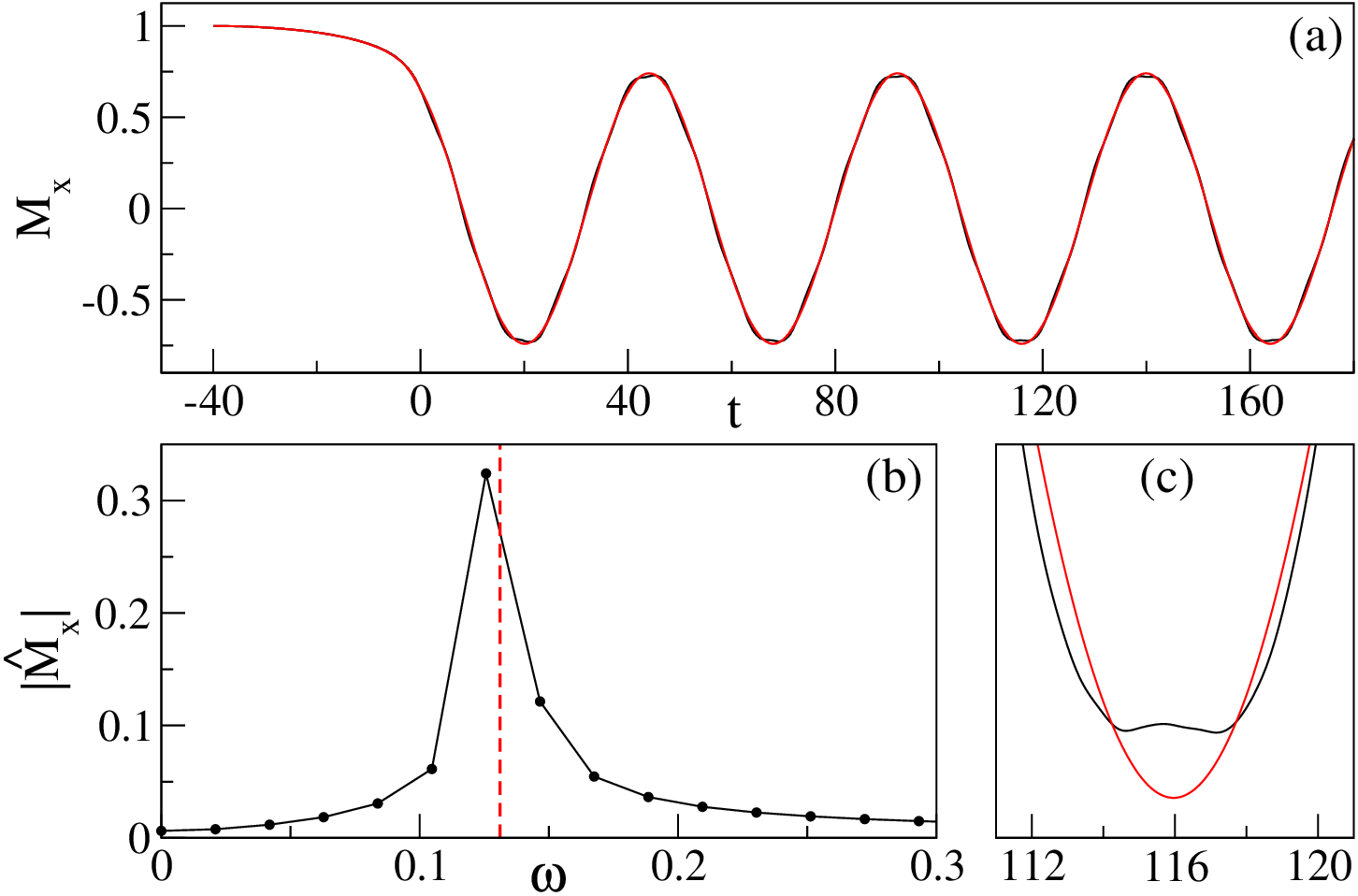}
\caption{Dynamics of the longitudinal magnetization in a periodic system for a nearly adiabatic 
quench that stops  at the critical point. 
This figure is organized in the same way as  Fig. \ref{per_0_85_fig}. 
The parameters 
are  $g_f=1$, $N=12$, $\tau_Q=40$, and $L=300$.
}
\label{per_1_fig}
\end{figure}

We will thus use the first above-mentioned approach, i.e., the direct numerical
solution, to characterize dynamics of the $N=12$  periodic chain. In the
next section, we will sacrifice the translational invariance by doing 
calculations in much larger  chains with open boundaries,
where such  problems with computation
of 
the longitudinal magnetization are absent.

\begin{figure}[t]
\includegraphics[width=\pref\textwidth,clip=true]{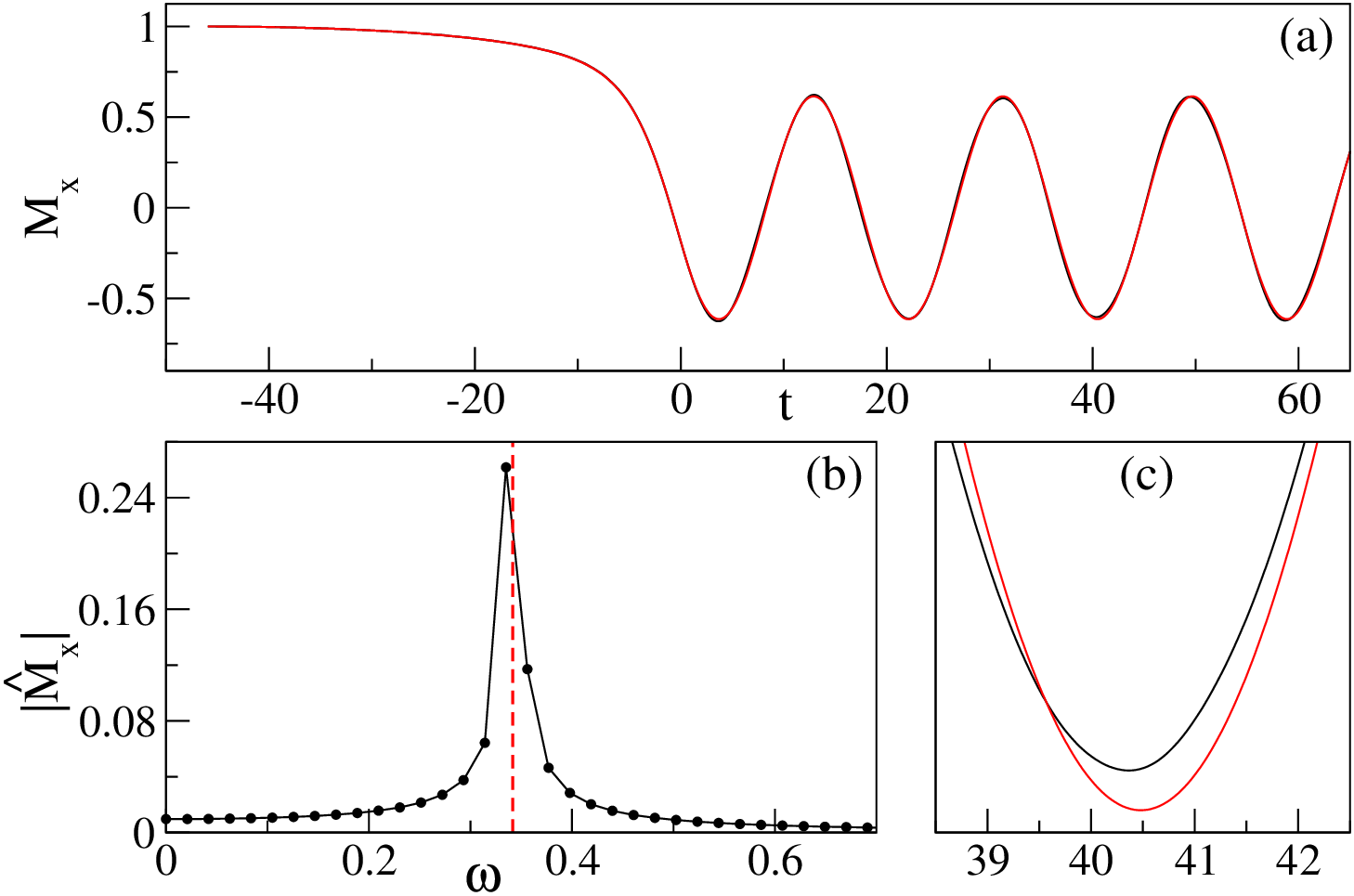}
\caption{Dynamics of the longitudinal magnetization in a periodic system for a nearly adiabatic 
quench that stops in the paramagnetic phase. 
This figure is organized in the same way as  Fig. \ref{per_0_85_fig}. 
The parameters 
are  $g_f=1.15$, $N=12$, $\tau_Q=40$, and $L=300$.
}
\label{per_1_15_fig1}
\end{figure}

The results of our numerical simulations, starting from initial state
(\ref{initial}) and employing quench protocol (\ref{godt}),  are presented in Figs.
\ref{per_0_85_fig}--\ref{per_1_15_fig1}, where we have chosen quench time $\tau_Q$
large-enough to ensure that  evolutions will be nearly adiabatic. We
see there that after stopping the driving, say for 
\be
0\le t\le L,
\ee
there are
periodic oscillations, which will be studied in  various ways. 

First, we will  extract  from numerical
data  the difference between  positions of the first two maxima of $M_x(t)$
and use it to compute   the frequency $\omega_{12}$ of oscillations. We
will also get  from such data  half of the difference between  $M_x(t)$ in
the first maximum and the second minimum to obtain the oscillation amplitude  $A_{12}$. 
These quantities are illustrated in Fig. \ref{per_0_85_fig}. 
The frequency $\omega_{12}$ and amplitude $A_{12}$
estimate $\delta(g_f)$ and $M_x^\eq(g_f)$--see (\ref{Mxeqq_bis})
for  justification of this statement\footnote{Note that we work with $\hbar=1$, where  the frequency and the
energy  can be directly compared.}.
As far as numerical simulations are concerned, the uncertainty  of getting $\omega_{12}$
and $A_{12}$ is negligible.
This is perhaps the easiest way of
characterization of  after-quench oscillations as it requires observation of no more than two oscillation periods.

Second, we will fit the periodic function
\be
M_x(t)=A_\fit\cos(\omega_\fit t+\text{const})
\label{fit}
\ee
expecting that $\omega_\fit$ and $A_\fit$ will estimate the same
physical quantities as $\omega_{12}$ and $A_{12}$.
Such fitting will be done with NonlinearModelFit function from \cite{Mathematica}. This 
function also provides  uncertainty of the fitted parameters, which is  negligible in 
our studies of  $\omega_\fit$ and $A_\fit$.
A few oscillation periods  provide  enough data for reaching
the full potential of this technique.

Third, we  will compute the discrete Fourier transform
\be
\begin{aligned}
&\hat M_x(\omega_j)=\frac{1}{S}\sum_{s=0}^{S-1}M_x(s\Delta t)e^{\ii \omega_j s\Delta t}, \\
&\Delta t=\frac{L}{S}, \ \omega_j=\frac{2\pi j}{L}, \  j=0,1,\dots,S-1,
\end{aligned}
\label{Fourier}
\ee
where $S$ is the number of data points that we generate during free evolution.
We will work with $S$ being   of the order of a few tens of thousands
(its exact value is of marginal use in the following discussion).
We will look for the global maximum of $|\hat M_x|$ studying  its position 
 $\omega_\text{max}$ and  value $|\hat M_x(\omega_\text{max})|$. 
 The former will estimate $\delta$. The latter,
 after multiplication  by a factor of two, will be of the order of  $M_x^\eq$. We will
 thus introduce $A_\text{max}=2|\hat M(\omega_\text{max})|$  in analogy to the notation that we have 
 used  above.
Both remarks  follow from (\ref{Mxeqq_bis}) and   (\ref{Fourier}).

We have applied all these techniques to numerics from Figs. 
\ref{per_0_85_fig}a--\ref{per_1_15_fig1}a getting results that are 
 presented in Table \ref{tab_dyn}. We see there   very good agreement between the oscillation
frequencies $\omega_{12}$ and  $\omega_\fit$ and the 
values of the symmetry-breaking gap $\delta$ from Table \ref{tab_eq}.
Moreover, similar  agreement is found between the oscillation amplitudes
$A_{12}$ and  $A_\text{fit}$ listed in Table \ref{tab_dyn} and $M_x^\eq$ from Table \ref{tab_eq}.

If we, however, use data from these  tables to compare  $\omega_\text{max}$ and
$A_\text{max}$ to the symmetry-breaking gap and the equilibrium longitudinal
magnetization, we will find much larger relative discrepancies.
Estimation of the symmetry-breaking gap through $\omega_\text{max}$ is mostly affected by 
 the spectral resolution of the Fourier transform  
\be
2\pi/L.
\label{spectral}
\ee
This is seen 
in  Figs.  \ref{per_0_85_fig}b--\ref{per_1_15_fig1}b, where the horizontal 
spacing between the data points ranges  between
$0.006$  and $0.02$, which is non-negligible  relative to $\delta$.
Estimation of the equilibrium longitudinal
magnetization through $A_\text{max}$ is additionally influenced by population of different Fourier
modes decreasing $|\hat M(\omega_\text{max})|$ with respect to its adiabatic value.
This can be seen through Plancherel
theorem. Finally,  mismatch between the symmetry-breaking gap and the
frequency grid $\{\omega_j\}$ additionally affects  $|\hat M(\omega_\text{max})|$.

A good agreement that we have found using the first two approaches is the
result of choosing so large  quench time $\tau_Q$, that  
evolutions are nearly adiabatic.
If we now  take a look at  
Figs. \ref{per_0_85_fig}c--\ref{per_1_15_fig1}c, we will be able to notice
small deviations from perfectly adiabatic solution  (\ref{Mxeqq}).
Similar features will be seen in the simulations of chains with open
boundaries, which
brings us to the next section of this work, where they will be discussed.

\begin{table}[t]
\begin{tabular}{ l | c | c| c| c| c | c}
           & $\omega_{12}$ & $\omega_\text{fit}$ &$\omega_\text{max}$ & $A_{12}$  & $A_\text{fit}$ & $A_\text{max}$  \\
\hline
$g_f=0.85$ & 0.028422 & 0.028747 & 0.031416            & 0.8620 & 0.8538  &  0.64   \\
\hline  
$g_f=1$    & 0.13114  & 0.13108  & 0.12566            & 0.7286  & 0.7397  &  0.65 \\
\hline
$g_f=1.15$ & 0.34177  & 0.34174   & 0.33510            & 0.6178 &  0.6146 &  0.52
\end{tabular}
\caption{Parameters describing free oscillations of the longitudinal magnetization from Figs.
\ref{per_0_85_fig} to \ref{per_1_15_fig1} (top to bottom). Symbols $\omega_{12}$, $\omega_\text{fit}$,$\dots$,
$A_\text{max}$  are defined around
equations (\ref{fit}) and (\ref{Fourier}). The fitting leading to
$\omega_\fit$ and $A_\fit$ has been done on time intervals
$0\le t \le 4\times2\pi/\delta(g_f)$ corresponding to four oscillation periods
during free evolution.
}
\label{tab_dyn}
\end{table}

\section{Dynamics of   chains with open boundaries}
\label{Open_sec}
We will study in this section dynamics of   Ising chains with open
boundaries. Such systems are described  by the following Hamiltonian
\be
 \tilde H(g) = -\sum_{i=1}^{N-1}\sigma^x_i\sigma^x_{i+1} - \sum_{i=1}^N g \sigma^z_i.
\label{OBC}%
\ee
There are at least two reasons for their consideration.

First, they are  more experimentally-relevant than the 
periodic chains  as it is a very complicated task to engineer a periodic
coupling between the spins  (see e.g. \cite{KorenblitNJP2012} for an
elaborate proposal how this might be achieved in a cold ion simulator of spin systems).

Second, computations of the longitudinal magnetization can be  efficiently done
in the  chains with open boundaries.
It is so because the parity operator (\ref{HP})
does not appear during the diagonalization 
of Hamiltonian  (\ref{OBC}) and so both parity subspaces are diagonalized 
with one and the same set of transformations 
(see Appendix \ref{Mx_app} for explanation why this is not the case in periodic
chains and 
note that $\tilde H$ commutes with the parity operator).
This allows  for efficient evaluation of the longitudinal magnetization in
systems composed of up to a few hundreds of spins, which is a major step
forward with respect to our studies of periodic chains.
This flexibility with respect to the system size allows us for systematic
studies of the transition from the non-equilibrium regime, where the
Kibble-Zurek theory describes the system's excitation, to the nearly  adiabatic regime, which we
have already begun to investigate in Sec. \ref{Periodic_sec}.

Moving on to the actual calculations, we mention that technical details
related to 
diagonalization of Hamiltonian (\ref{OBC}) and time evolution that it generates are
presented in Appendix \ref{OBC_app}. We only summarize  here  some basic
formulae that are necessary for understanding of the following discussion.

After Jordan-Wigner mapping (\ref{JW}), the Hamiltonian is
diagonalized by the Bogolubov transformation so that it finally reads
\be
\begin{aligned}
&\tilde H = \sum_{i=1}^N E_i \B{\gamma_i^{\dagger} \gamma_i - 1/2},\\
& \lbrace \gamma_i, \gamma_j^{\dagger} \rbrace = \delta_{ij}, \ \lbrace
\gamma_i, \gamma_j \rbrace = 0,
\end{aligned}
\label{Hbog}
\ee
where the energies of single-particle excitations, defined in our work  with respect to
the ground state energy, are sorted   in ascending order,
$E_1 \leq E_2 \leq \dots \leq E_N$,  so that 
the symmetry-breaking
gap is
\begin{equation}
\delta=E_1.
\label{dE1}
\end{equation}

They are given by 
\begin{equation}
E_i = 2\sqrt{g^2-2g\cos\theta_i + 1},
\label{Ei}
\end{equation}
where  $\theta_i$'s are obtained from \cite{CabreraPRB1987}
\begin{equation}
g\sin\BB{(N+1) \theta} = \sin(N \theta), \quad 0<\theta < \pi.
\label{pppp}
\end{equation}

For $g>N/(N+1)$, $\theta_1,\dots,\theta_N$ are the  real roots of
(\ref{pppp}).
For $0\le g \le N/(N+1)$, most interestingly, there is one purely imaginary solution of (\ref{pppp}),
$\text{Re}(\theta_1)=0$. Besides that,
there are $N-1$ real roots of (\ref{pppp}): $\theta_2,\dots,\theta_N$.
Equation (\ref{pppp}) cannot be solved analytically for an arbitrary value of
the magnetic field $g$. Its numerical solutions, relevant for the subsequent discussion, are
collected in Table \ref{tab_Eobc}. 

\begin{table}
\begin{tabular}{ l | c | c | c | c | c | c | c}
  &  $E_1$ & $E_2$ & $E_3$ & $E_4$ & $E_5$ & $E_6$ & $E_7$  \\
  \hline
  $g_f=0.85$, $N=20$ &  0.0217       &   0.470   &   0.730 & & &  	 \\
  \hline
  $g_f=1$, $N=50$ &  0.0622 & 0.187 & 0.311 & 0.435 & 0.558 & 0.681 & 0.803	 \\
  \hline
  $g_f=1.15$, $N=50$ & 0.322  &  0.383  	 &  0.470 & & &
\end{tabular}
\caption{Energies of single-particle  excitations
relevant for deciphering the positions of  marked maxima in Figs. \ref{fer_open_fig}, \ref{para_open_fig}, and
\ref{fat_fig}.}
\label{tab_Eobc}
\end{table}
\begin{table}[t]
\begin{tabular}{ l | c }
           & $M_x^\eq$  \\
\hline
$g=0.85$, $N=20$ &  0.8448 \\
\hline
$g=1$, $N=50$  & 0.6188  \\
\hline  
$g=1.15$, $N=50$ & 0.4204
\end{tabular}
\caption{The equilibrium longitudinal magnetization in the Ising chain with open boundaries 
for the parameters 
relevant to the studies reported in Figs. \ref{fer_open_fig}--\ref{many_fig}. 
}
\label{Mxopen_tab}
\end{table}

Besides the symmetry-breaking gap, we are also interested in the longitudinal
magnetization, which is now position dependent. Its  equilibrium value in the  Ising chain with open 
boundaries  was 
recently discussed in \cite{Oskar}, where it was analyzed how 
the ``ends'' of the chain affect its value. To minimize
their influence on our results, we will focus our attention on the center of the system
by calculating 
\be
M_x=\la\sigma^x_{N/2}\ra.
\label{Mx_open}
\ee
The  equilibrium values of such defined  longitudinal magnetization, for the parameters relevant for the subsequent
discussion, are listed in Table \ref{Mxopen_tab}.
The technical details of
computation of (\ref{Mx_open}) are discussed in Appendix \ref{OBC_app}.

An important thing now is to note that if we write the  Schr\"odinger-picture
wave-function at the time the quench stops as 
\be
\begin{aligned}
&|\psi(t=0)\ra=\sum_\idots a_\idots|\idots\ra,\\
&\sum_\idots |a_\idots|^2=1,\\
&|\idots\ra=\B{\gamma_1^\dag}^{i_1}
\B{\gamma_2^\dag}^{i_2}\cdots\B{\gamma_N^\dag}^{i_N}
|GS\ra,\\
&i_n=0,1 \ \text{and} \   \gamma_n|GS\ra=0 \for 1\le n\le N,
\end{aligned}
\ee
then at times $t>0$    
\be
M_x(t)=\sum_{\overset{\jdots}{\idots}}  
\overline{a_\jdots}
a_\idots  e^{\ii t\sum_{n=1}^N(j_n - i_n)E_n}
\la\jdots|\sigma^x_{N/2}|\idots\ra
\label{rrttyy}
\ee
with
\be
\sum_{n=1}^N (i_n+j_n) \ \text{being  odd}.
\label{rrttyy1}
\ee
If this condition  is not satisfied, then the matrix element in (\ref{rrttyy}) identically  vanishes,
which can be shown with Wick's theorem. This feature impacts Fourier spectra
of the longitudinal magnetization.

\begin{figure}[t]
\includegraphics[width=\pref\textwidth,clip=true]{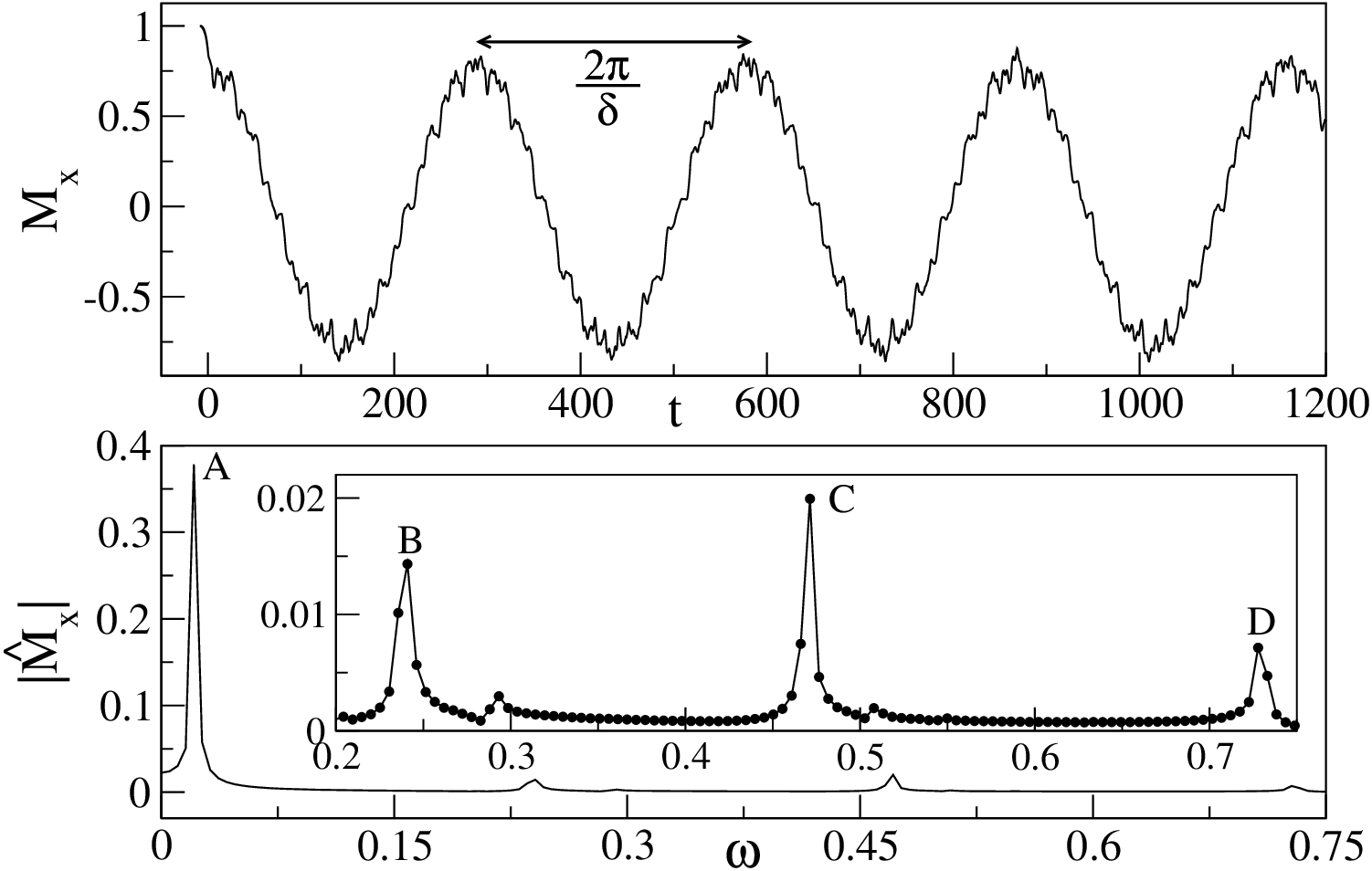}
\caption{
Upper panel: dynamics of the longitudinal magnetization in the  Ising
chain with open boundaries for the quench
that  stops in the ferromagnetic phase. The parameters are 
$g_f=0.85$, $N=20$,  and  $\tau_Q=10$. $\delta$ is  given by $E_1$ from
Table \ref{tab_Eobc} (first row).  
Lower panel: modulus of the discrete Fourier transform of free evolution
data, i.e. $M_x(t>0)$,  from the upper panel.
The inset enlarges peaks B--D.
$L=1200$ has been used to compute the  transform. 
}
\label{fer_open_fig}
\end{figure}
\subsection{Quenches stopping in  ferromagnetic or paramagnetic phase}
\label{QFP}
We start discussion of our numerical simulations  from    
the non-equilibrium quench that has been stopped  in   the 
ferromagnetic  phase (Fig. \ref{fer_open_fig}).
The first thing that catches our attention in this figure is ``roughness'' of free evolution of $M_x$.
To understand it, we need to  look at modulus of its  discrete Fourier transform. 

By doing so, we first notice   a series of peaks enumerated by $A$, $B$, etc. whose
maxima are  placed at $\omega_A$, $\omega_B$, etc. listed in Table
\ref{omegi_tab}. If we now use data from  Table \ref{tab_Eobc} and equations (\ref{rrttyy}) and (\ref{rrttyy1}),
we can note that within spectral resolution of the Fourier transform  $\omega_A$, $\omega_B$, $\omega_C$, and $\omega_D$
can be identified with $E_1$, $E_3-E_1-E_2$, $E_2$, and $E_3$, respectively. 

Occupation of the Fourier modes around $\omega_A$ is responsible for the 
oscillation period  $2\pi/\delta(g_f)$  marked in the upper panel
of Fig. \ref{fer_open_fig}.
Next, we note that (i)
$\omega_B$, $\omega_C$, etc. are larger  by at least a factor of ten   than $\omega_A$
and (ii) $|\hat M_x(\omega_A)|$ is larger by at least a factor of fifteen 
than $|\hat M_x(\omega_B)|,|\hat M_x(\omega_C)|$, etc.
The (i) observation means that there will be high frequency oscillations on
top of the base  oscillation, whose frequency is approximated by $\omega_A$. 
The (ii) remark implies that they will
have small amplitude relative to the amplitude of the base  oscillation. 
Both features are nicely seen in the upper panel of Fig.
\ref{fer_open_fig}.
They explain small fluctuations of the data presented there.

\begin{figure}[t]
\includegraphics[width=\pref\textwidth,clip=true]{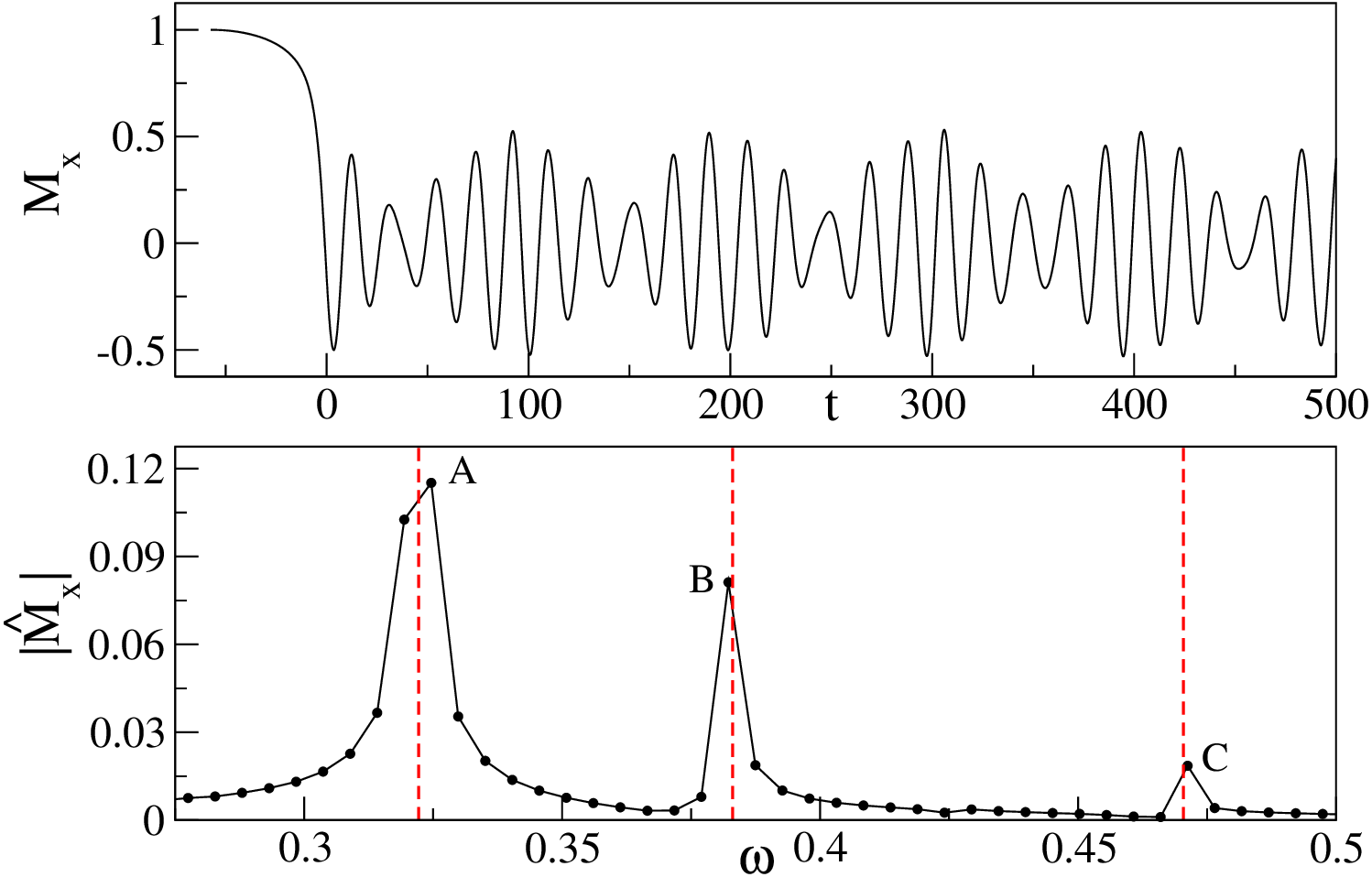}
\caption{Upper panel: dynamics of the longitudinal magnetization in the Ising
chain with open boundaries for the quench that  stops in the paramagnetic  phase. The parameters are 
$g_f=1.15$, $N=50$, and $\tau_Q=50$.  
Lower panel: modulus of the discrete Fourier transform of the free evolution data from the 
the upper panel.
The vertical red dashed lines in the lower panel show energies of the
single-particle excitations from the third row of Table \ref{tab_Eobc}. 
$L=1200$ has been used to compute the  transform. 
}
\label{para_open_fig}
\end{figure}

A quite different situation is encountered when the quench stops in the
paramagnetic phase. This is illustrated in Fig. \ref{para_open_fig}, where we see beats.
A simple explanation of this observation comes again from the discrete Fourier
transform, where we see  two leading peaks centered around $\omega_A$ and $\omega_B$.
The beats result from the fact that $|\hat M_x(\omega_A)|$ is comparable
to $|\hat M_x(\omega_B)|$. Moreover, we note that $\omega_A$, $\omega_B$, and
$\omega_C$ are of the same order of magnitude, which eliminates high frequency oscillations
seen in Fig. \ref{fer_open_fig}. Looking more quantitatively at the Fourier
transform from Fig. \ref{para_open_fig}, we notice that
$\omega_A$, $\omega_B$, and $\omega_C$ can be identified with
$E_1$, $E_2$, and $E_3$ within the spectral resolution of the Fourier
transform (Tables \ref{tab_Eobc} and \ref{omegi_tab}).

Finally, at the risk of stating the obvious, we mention that 
 we recover adiabatic results, akin to 
those presented in Figs. \ref{per_0_85_fig} and \ref{per_1_15_fig1}, by 
increasing  the quench times from Figs. \ref{fer_open_fig} and
\ref{para_open_fig}.
In the opposite limit of fast transitions, free evolution of the longitudinal
magnetization becomes noisy and so less  interesting in the context of our  studies.

\subsection{Quenches to critical point}
\label{QCP}
\begin{figure}[t]
\includegraphics[width=\pref\textwidth,clip=true]{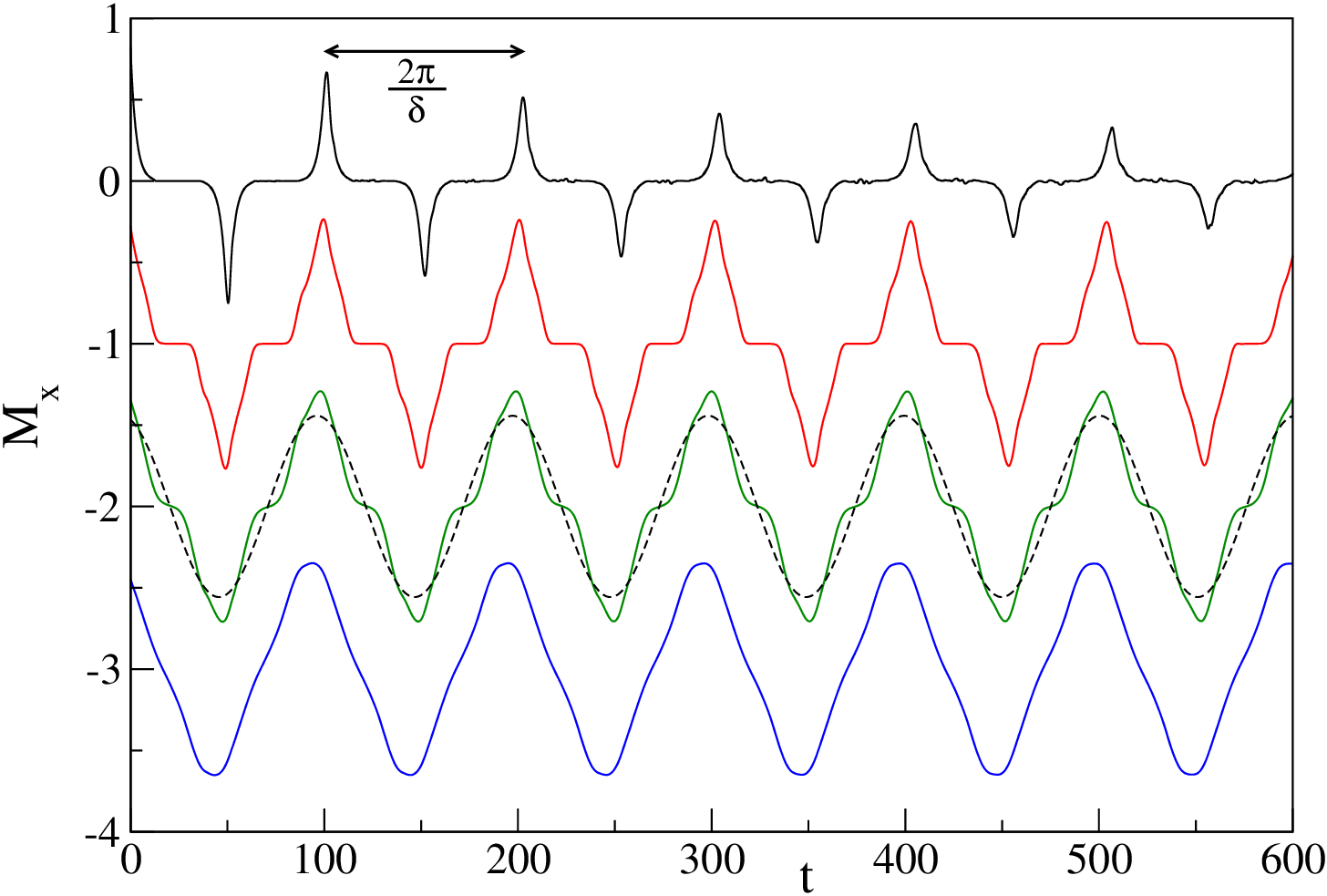}
\caption{Free evolution of the longitudinal magnetization after 
quenches to the critical point in the Ising chain with open boundaries. 
The solid curves, from top to bottom,
correspond to $\tau_Q=2$ (black), $\tau_Q=20$ (red), $\tau_Q=73$ (green), and
$\tau_Q=200$ (blue). 
The subsequent curves are shifted downward by a multiple of $1$ to
facilitate their comparison.
The dashed black  curve shows
$0.557\cos(0.0622t+0.307)$ shifted  downward by $2$. It comes from
the fit to the numerics for $\tau_Q=73$, which is close to the crossover
quench time discussed in Fig. \ref{dev_fig}.
The parameters are $g_f=1$ and $N=50$. $\delta$ is given by $E_1$ from Table \ref{tab_Eobc} (second row).
}
\label{many_fig}
\end{figure}

We will discuss now the transition to the adiabatic regime for evolutions ending
 at the critical point. Such evolutions are depicted in Fig. \ref{many_fig}.
 For small $\tau_Q$, we see a train of narrow peaks, whose amplitude
  decreases as time goes by. The  magnetization in
 between the peaks is nearly zero. 
As  evolutions slow down, the decay of the peaks' amplitude slows and the peaks' width increases shrinking  time
intervals, where the system is  unmagnetized in the longitudinal direction. By the time those intervals
disappear, the curve describing dynamics of the longitudinal magnetization has a triangular-like shape.
Further increase of the quench time brings the expected single-frequency
dynamics characteristic of the adiabatic evolution (\ref{Mxeqq_bis}).

\begin{figure}[t]
\includegraphics[width=\pref\textwidth,clip=true]{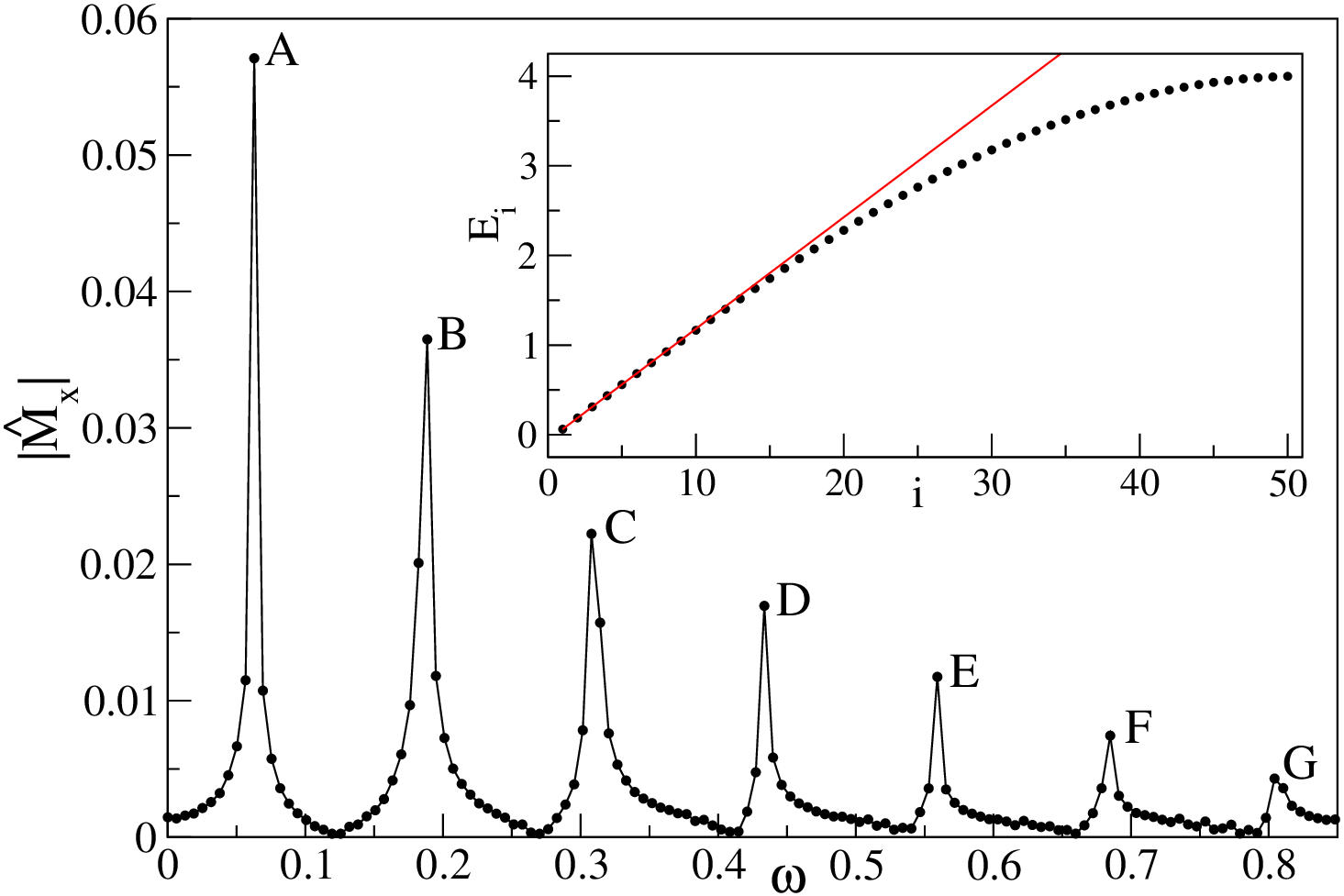}
\caption{Modulus of the discrete Fourier transform of  free evolution 
of the longitudinal magnetization of the fastest quench from Fig.
\ref{many_fig}: $g_f=1$, $N=50$, $\tau_Q=2$. $L=1000$ is used to
compute the transform. The inset shows (\ref{Ei})  for $g=1$ and $N=50$. 
The black dots in the inset are obtained from numerical
solution of   (\ref{pppp}),
while the solid red line  shows (\ref{Ei2i}) with $\delta(g=1)$ given
exactly by (\ref{gapp}). 
}
\label{fat_fig}
\end{figure}

The rather unusual shape of  oscillations of the longitudinal magnetization for the fastest
quenches from Fig. \ref{many_fig} comes from substantial population of several Fourier
modes. This is illustrated in Fig. \ref{fat_fig}, where the subsequent Fourier
peaks are centered at the energies of consecutive single-particle excitations (Tables
\ref{tab_Eobc} and \ref{omegi_tab}).

The question now is why the period of these 
oscillations is approximately given by $2\pi/\delta(g=1)$.
This would be an expected result for  adiabatic  evolutions, where the
oscillation pattern would be cosinusoidal (\ref{Mxeqq_bis}). It may  thus
be a bit surprising that multi-frequency  oscillations from Fig. \ref{many_fig}
exhibit the same period. This can be understood 
by noting that 
\be
E_i\approx (2i-1)\delta(g=1)\approx(2i-1)\frac{\pi}{N}
\label{Ei2i}
\ee
for the lowest-energy modes--see the second row of Table \ref{tab_Eobc}, 
(\ref{gapp}), and the inset of Fig. \ref{fat_fig}. 
If such a relation would hold for
all $E_i$'s, then the oscillations of the longitudinal magnetization during
free evolution would be perfectly adiabatic with the $2N$ period regardless 
of the quench time $\tau_Q$. 
Relation (\ref{Ei2i}), however, is approximate, which we also illustrate 
in the inset of Fig. \ref{fat_fig}. This  explains quasi-periodicity
of the oscillation pattern for the fastest quench in Fig. \ref{many_fig}.
For a bit slower, but still non-adiabatic quenches depicted in this figure,
(\ref{Ei2i})  properly captures these single-particle
excitation energies that give the main contribution to the free dynamics of
$M_x(t)$. This  is sufficient for explanation of the oscillation period from
Fig. \ref{many_fig}.

This  oscillation period 
is twice smaller than the oscillation  period at the critical point
of the periodic system, which can be trivially   shown with
(\ref{gapcritical}).
The linear dependence of the oscillation period 
on the system size, albeit with a different prefactor,  was also observed in our earlier studies 
 of the quantum Ising model \cite{MichalOne}, where we investigated  free
dynamics of the transverse magnetization  after  quenches 
moving  the system from the paramagnetic phase to the critical point. 
Finally, we mention  that approximation  (\ref{Ei2i}) works  only at  the critical
point, which explains why different dynamics have been observed in Sec. \ref{QFP}.

To study quantitatively the crossover  from the non-equilibrium to the adiabatic regime, 
which we depict in Fig. \ref{many_fig},
we
need some measure of the deviation of non-equilibrium evolution from the
adiabatic limit. A good measure should be easily numerically and
experimentally accessible. It should be also stable against fluctuations of
the data for $M_x(t)$. Several options seem to be  available.

\begin{table}[t]
\begin{tabular}{ l | c | c| c| c | c | c | c }
           & $\omega_A$ & $\omega_B$ & $\omega_C$ & $\omega_D$  & $\omega_E$ & $\omega_F$ & $\omega_G$ \\
\hline
$g_f=0.85$, $N=20$ &  0.0209   & 0.241      &  0.471    & 0.728 & &  \\
\hline  
$g_f=1$, $N=50$    & 0.0628 & 0.188 & 0.308 & 0.434 & 0.559 & 0.685 & 0.804 \\
\hline
$g_f=1.15$, $N=50$ & 0.325    &  0.382     &  0.471   & & & &
\end{tabular}
\caption{The positions  of maxima of $|\hat M_x|$ marked in  Figs. \ref{fer_open_fig}, \ref{para_open_fig},  and \ref{fat_fig}. 
}
\label{omegi_tab}
\end{table}

First, one may  analyze  modulus of the discrete Fourier transform. For example, one can study 
how the global maximum around the symmetry-breaking gap grows with increasing $\tau_Q$.
Alternatively, one may  research  how  the other
extrema disappear in such a limit. This choice, however, is problematic for
the reasons explained by the end of Sec. \ref{Periodic_sec}. 
For example, there are limitations imposed by the  spectral resolution of the discrete Fourier transform 
(\ref{spectral}). To overcome them, either long free evolution times are needed or
some fitting procedure allowing for precise interpolation of the properties of
extrema of $|\hat M_x|$ from sparse data. This is a
complication affecting both numerical and experimental studies. The latter
would be also affected by the fact that the Fourier transform is not directly
measured and so its extraction out of $M_x(t)$
will necessarily
bring some inaccuracies that may play a role in the Kibble-Zurek scaling analysis.

Second, one may use  a more straightforward approach by studying the amplitude  and
spacing  of the first two peaks of $M_x(t)$, just as in Sec. \ref{Periodic_sec}.
 Such a method, however, is
susceptible to fluctuations of the data. This can be improved by averaging results 
collected for several peaks, but this would again require long free evolution
times, which is problematic.

\begin{figure}[t]
\includegraphics[width=\pref\textwidth,clip=true]{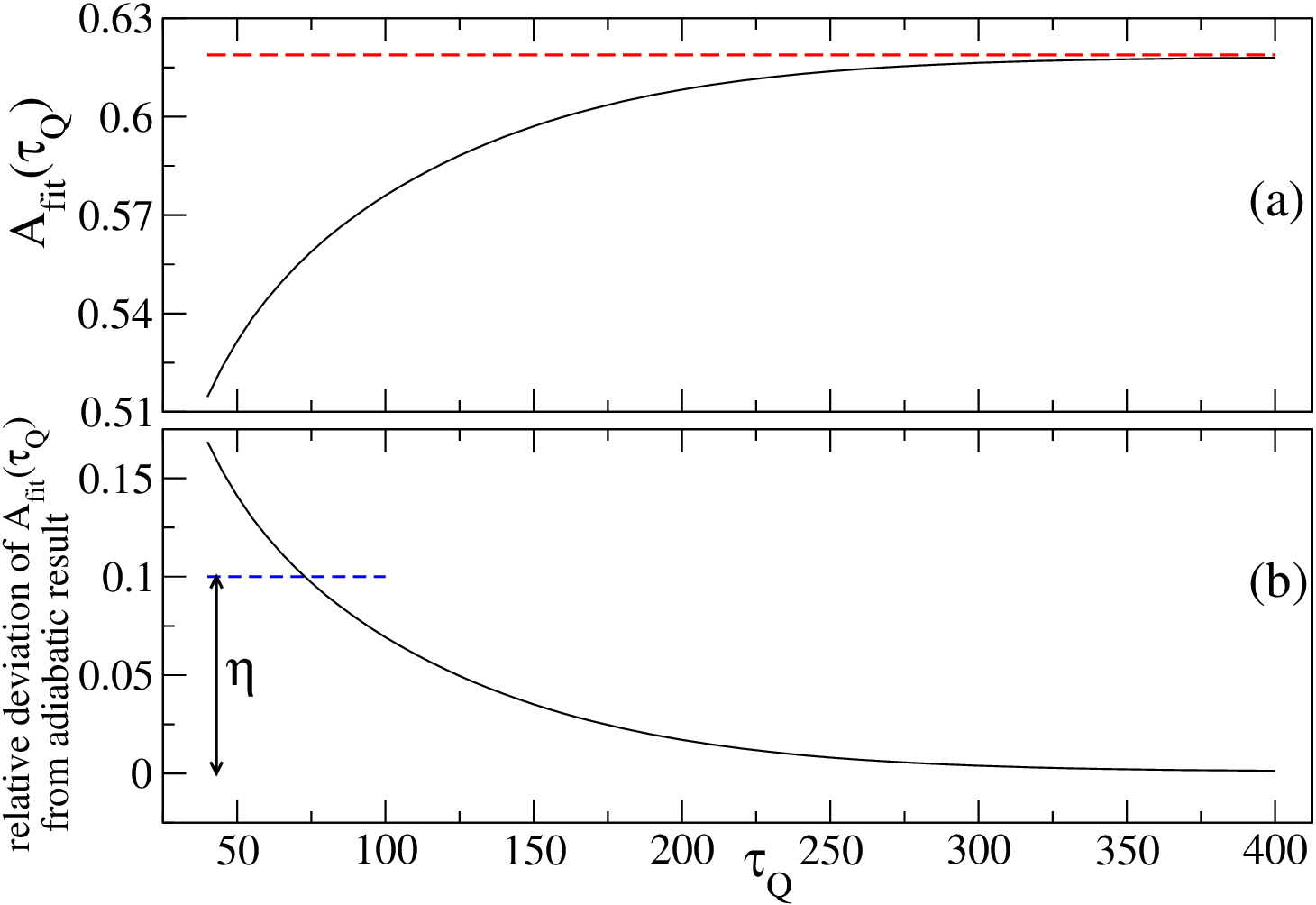}
\caption{Panel (a): the fitted amplitude of oscillations of the longitudinal magnetization
after stopping the quench at the critical point of the  Ising chain with open boundaries. 
The solid black line  shows numerics in both panels.
The dashed red line shows
the result for perfectly adiabatic evolution to the critical point,
which is given by the equilibrium magnetization (second row of Table
\ref{Mxopen_tab}).  Panel (b): the left-hand side of
(\ref{tcross}). 
The dashed blue line shows the threshold $\eta=10\%$ from  
(\ref{tcross}). Its intersection with the solid black line gives
$\tau_Q^\cross\approx72.83$. The  system size is $N=50$. 
}
\label{dev_fig}
\end{figure}

Third, one may fit  (\ref{fit}) to  free evolution of the longitudinal magnetization and study 
such obtained amplitude of oscillations $A_\fit$ (Fig. \ref{dev_fig}).  
Such a procedure uses all information contained in $M_x(t)$--not only the one stored 
in the extrema of either $M_x$ or $|\hat M_x|$--and so long evolution times are not needed.
Moreover, it   should work well with irregular
data averaging out the fluctuations, which is of interest in the context of high-precision 
numerical and experimental research. This is the approach that we will employ.

Before moving on, we again mention that the fitted amplitude $A_\fit$  converges to 
$M^\eq_x(g_f)$ in the adiabatic limit.
For faster quenches, however, it  underestimates the real amplitude of oscillations,
which is seen  in Fig. \ref{many_fig}. This has no effect on
our studies, which is perhaps best illustrated by the excellent agreement 
between the scaling exponent extracted out of the fitted amplitude and the
predictions of the Kibble-Zurek theory (see below).

To proceed, we define the crossover quench time $\tau_Q^\cross$ by the condition
\be
\left|\frac{A_\fit(\tau_Q)-M^\eq_x(g_f)}{M^\eq_x(g_f)}\right|< \eta \ 
\for \ \tau_Q>\tau_Q^\cross,
\label{tcross}
\ee
where $\eta$ is the threshold set on  the relative  
difference between the fitted amplitude of oscillations and its asymptotic in $\tau_Q$ value.
We will use in this formula the amplitude obtained by fitting (\ref{fit}) to $M_x(t)$ for $0\le t\le12N$,
which corresponds to roughly $6$ oscillation periods in the  chain with open
boundaries. 
Moreover, we will set $\eta=10\%$, which should be large-enough to be
experimentally-relevant and small-enough to describe the crossover to the adiabatic limit.

Our results for $A_\fit$, in the experimentally-relevant system composed of $N=50$ spins 
\cite{LukinNature2017,MonroeNature2017,LukinNature2019}, are presented in 
Fig. \ref{dev_fig}a, where we see that the fitting procedure produces a
perfectly smooth curve monotonically approaching $M^\eq_x(g_f)$. The threshold
$\eta$  is illustrated in Fig. \ref{dev_fig}b.

\begin{figure}[t]
\includegraphics[width=\pref\textwidth,clip=true]{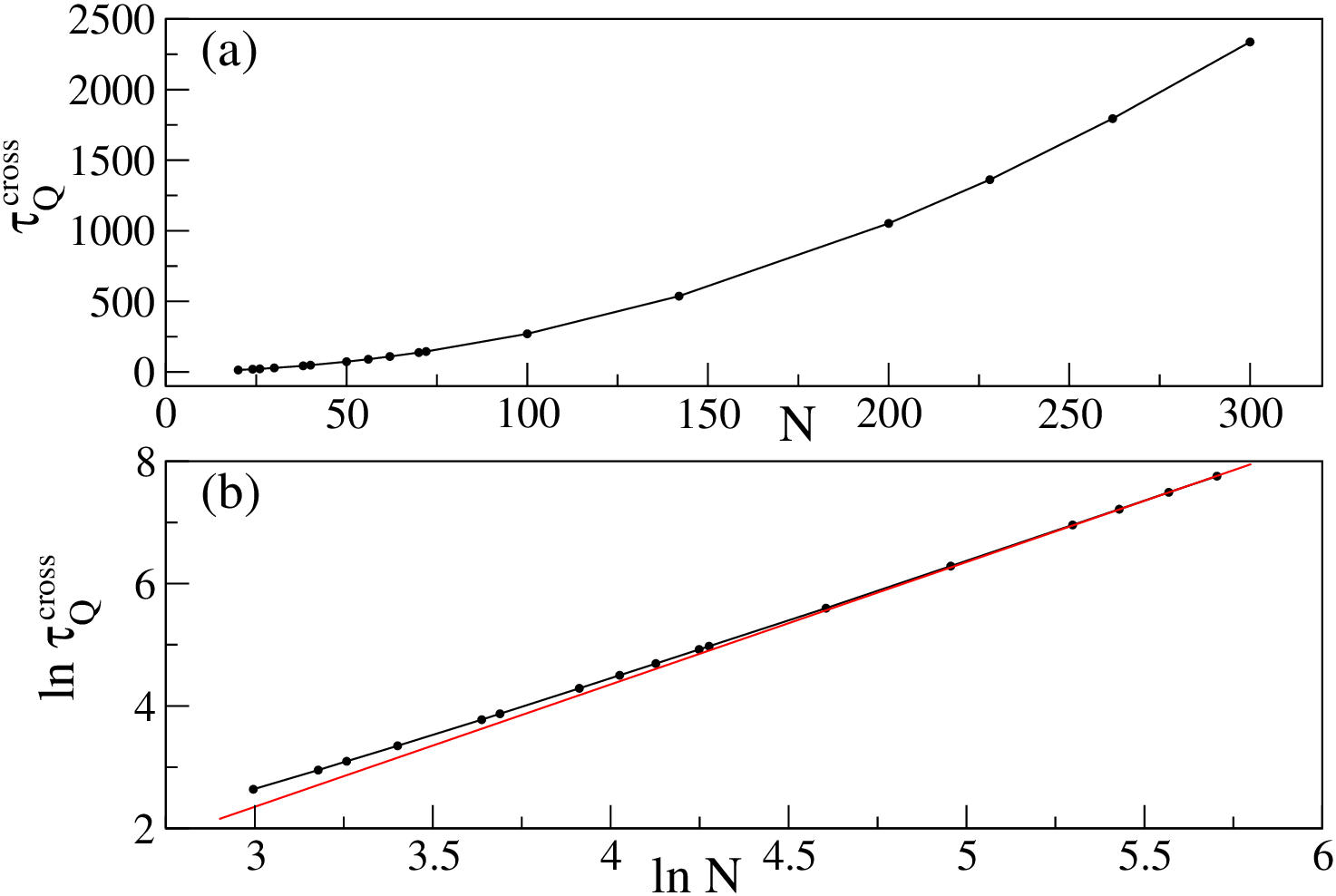}
\caption{Panel (a): the crossover quench time $\tau_Q^\cross$ as a function of the system
size for quenches that stop at the critical point of the Ising chain with open boundaries
($g_f=1$). Dots come from numerics,
 line segments join them to guide the eye. 
The threshold from (\ref{tcross}) is  $\eta=10\%$.
Panel (b): data from the upper panel shown on a double logarithmic plot. The solid red line 
has the slope equal to $+2$ and the intercept fitted to the two
largest-$\tau_Q$ data points. 
}
\label{kz_fig}
\end{figure}

Repeating such analysis for system sizes $20\le N\le 300$, we have obtained detailed
results for the crossover quench time $\tau_Q^\cross$, which we present in Fig.
\ref{kz_fig}a. As we anticipate   from (\ref{kzscaling}) that 
\be
\tau_Q^\cross(N)\sim N^a,
\ee
where $a>0$ is the scaling exponent, we  display results for $\tau_Q^\cross$ on a double logarithmic plot in Fig.
\ref{kz_fig}b.
Instead of  a straight line, we find in this figure a curve slightly bending upwards as the system size grows.
This means that the  exponent $a$  increases with $N$. To quantify this observation, we fit
\be
\ln\tau_Q^\cross = a\ln N + b
\label{cross_fit}
\ee
to numerical data from four  different ranges of the system sizes. The results are collected in Table \ref{kz_tab},
where we see that $a$ approaches the value of $2$ for the largest system sizes that we consider. 
This is in a very good agreement with the Kibble-Zurek
scaling argument (\ref{kzscaling}), which is supposed to work best in 
the large-system limit. Finally, we notice that the increase of $a$ with $N$ is monotonic, leaving no
doubts about  stability of the procedure of extraction of $\tau_Q^\cross$ from
the free evolution data for the longitudinal magnetization.

\section{Discussion}
\label{Discussion_sec}

The goal of this work was  to investigate how the symmetry-breaking gap,
which is of crucial importance in the discussion of quantum phase transitions, 
can be studied with the help of quantum quenches.
We have chosen for this purpose an exactly solvable model, the quantum Ising model in the transverse field,
and analyzed   its dynamics  after  quenches
induced by the gradual change of the magnetic field. These  quenches
start from the easy-to-prepare broken-symmetry ground state at the vanishing magnetic field.
They  bring the
system to the desired value of this field, where the symmetry-breaking gap
can be  read from the  subsequent   free-evolution dynamics of the longitudinal
magnetization. In this way a small symmetry-breaking gap can be seen through 
large-amplitude oscillations of the longitudinal magnetization.

We have discussed different ways of analyzing the oscillatory dynamics of such 
magnetization showing that one can also extract the 
equilibrium longitudinal magnetization out of them. All this can be accurately
done if the quench is slow enough, which we have  studied
in the context of the Kibble-Zurek theory
of non-equilibrium phase transitions. An excellent agreement between
predictions of this theory and the dynamics of the longitudinal magnetization
has been found. 

\begin{table}[t]
\begin{tabular}{ l | c | c }
  &  $a$ & $b$ \\
  \hline
  $20\le N \le 30$ &   1.75(1)  & -2.60(4)    	 \\
  \hline  
  $38\le N\le62$ &    1.871(3) & -3.03(1)    \\
  \hline  
  $70\le N\le142$ &    1.92(1) & -3.26(6)    \\
  \hline
  $200\le N\le300$ & 1.970(4) & -3.48(2)
\end{tabular}
\caption{The results of fitting (\ref{cross_fit}) to numerical data in
different ranges of the system sizes (either four or five data points are used for each linear regression).
We provide one standard error in the brackets delivered by the LinearModelFit
function from \cite{Mathematica}. 
}
\label{kz_tab}
\end{table}

Although our studies have been done in the Ising model, they can be
extended to other systems exhibiting the symmetry-breaking phenomenon. 
For example, the Ising-like ones  with long-range interactions 
that are typically found in cold ion and atom  emulators of spin systems
(see e.g. \cite{PorrasPRL2004,KorenblitNJP2012,LewensteinAdv,BlochSci2017,SchaussQST2018}). 
These systems provide a promising platform for experimental realization of the
studies discussed in our work for two reasons.
First, their size is  finite  rather than thermodynamic making  their 
symmetry-breaking gap large-enough to be experimentally measurable. 
This should not be taken for granted because  it is not the case in traditional
condensed matter setups discussed in the context of phase transitions. 
Second,  there has been 
substantial   progress in the experimental studies of the dynamics  of 
such systems (see e.g. \cite{LukinNature2017,MonroeNature2017,LukinNature2019}). 

Another promising platform for
implementation of our ideas is provided by D-Wave machines, which can be also
used for simulations of   spin models (see e.g. \cite{HarrisSci2018}). 
Quite interestingly, 
D-Wave-based investigations of non-equilibrium Kibble-Zurek dynamics of the one-dimensional quantum Ising
model in the transverse field were  recently reported  in \cite{JacekSciRep2018}.
It should be thus possible to use  our predictions  for critical assessment 
of the performance of such devices.

Talking about experimental realizations, the symmetry-breaking phenomenon
was recently experimentally investigated in \cite{LMGexp}.
These studies were done in a cold atom cloud, where each atom was simulating 
the $16$-spin Lipkin-Meshkov-Glick model. This is the Ising-like model
with the  nearest-neighbor spin-spin interactions replaced with identical couplings 
between all the spins. Besides exploration of a different Hamiltonian,
these interesting studies differ from our work  in 
the following aspects. First, the quenches   
start in the paramagnetic phase. Second, the initial state for
them occupies one of the parity subspaces  and so the symmetry-breaking 
perturbation is used to populate the other one as well. Third, perhaps most importantly, 
they are limited to one, rather small system size, and they do not explore 
the non-adiabatic Kibble-Zurek dynamics of the longitudinal magnetization,
which is of considerable  importance in our work.

Finally,   we mention that we  hope that this work will trigger interest in 
the experimental studies of the symmetry-breaking phenomenon, which could lead
to  quantitative insights into the very nature of quantum phase transitions.
This would be most interesting 
in systems  that can be neither analytically nor numerically studied  in the foreseeable
future.

\section*{Acknowledgments}
We thank Marek Rams for  useful discussions, sharing with us the results of
his numerical simulations,  reading the manuscript, and providing us with most
useful feedback.
MB and BD were supported by the Polish National Science Centre (NCN) grant
DEC-2016/23/B/ST3/01152. 

\appendix

\section{Longitudinal magnetization in Jordan-Wigner-transformed   periodic Ising chains}
\label{Mx_app}
We will explain here  the source of difficulties in computation of the longitudinal magnetization 
in the free fermion  representation of the periodic Ising model.
While doing so, we will also derive expressions for dynamical gaps
$\Delta_\pm$, which have been introduced in Sec. \ref{Idea_sec}.

Such a representation  comes from  employment of the Jordan-Wigner transformation
\ba
&\sigma^z_i=1-2 c_i^\dag c_i, \  \sigma^x_i=(c_i+c_i^\dag)\prod_{j<i}\B{1-2 c_j^\dag c_j},\\
&\{c_i, c_j^\dag\}=\delta_{ij}, \ \{c_i, c_j\}=0,
\label{JW}
\ea
after which  Hamiltonian (\ref{PBC}) reads
\be
 H(g) = -\sum_{i=1}^{N-1}  f_{i,i+1} - g \sum_{i=1}^N (c_i  c_i^\dag  -  c_i^\dag  c_i)+ f_{N,1} P,
\  f_{i,j} =  c_i^\dag c_j -  c_i  c_j^\dag -  c_i c_j +   c_{i}^\dag  c_j^\dag.
\label{zxcvbn}
\ee
The complication one encounters now is that the parity operator  is not quadratic in 
fermionic $c_i$ and $c_i^\dag$ operators. This  can be found by combining
(\ref{HP}) with (\ref{JW}). As a result, Hamiltonian (\ref{zxcvbn})
is non-quadratic. Its exact analytical diagonalization  is still possible, but 
the price one has to pay is that one has  to split the Hilbert space into
positive and negative parity subspaces imposing different boundary conditions
on the fermionic  operators in those subspaces \cite{BDJPA2014}. 
This is realized by going to the momentum space 
\be
c_j =\frac{e^{-\ii\pi/4}}{\sqrt{N}}\sum_k c_k e^{\ii kj},
\label{Fou}
\ee
and choosing different quantization 
schemes for momenta in the positive and negative parity subspaces, 
\be
k_+=\pm\frac{\pi}{N},\pm\frac{3\pi}{N},\dots,\pm\frac{N-1}{N}\pi, 
\label{kp}
\ee
and 
\be
k_-=0,\pm\frac{2\pi}{N},\pm\frac{4\pi}{N},\dots,\pm\frac{N-2}{N}\pi,\pi,
\label{km}
\ee
respectively. These expressions are valid for even $N$ 
(see \cite{BDJPA2014} for comprehensive discussion of the Ising
diagonalization intricacies). 
Therefore, different transformations are used to diagonalize 
the two parity subspaces in periodic chains. No such complications 
appear in the chains with open boundaries, where the 
parity operator is absent  in  Jordan-Winger-transformed Hamiltonian
(\ref{OBCnext}).

Combining the results of 
 \cite{JacekPRL2005}
and
\cite{BDJPA2014}, time-dependent wave-function (\ref{psit}) can be  obtained from
\begin{align}
\label{psipp}
&|\psi_+(t)\rangle =\prod_{k_+>0} \B{u_{k_+}(t) - v_{k_+}(t)
c_{k_+}^\dag c_{-k_+}^\dag}|{\rm vac}\rangle,\\
&|\psi_-(t)\rangle =e^{2\ii t}c_0^\dag\prod_{0<k_-<\pi} \B{u_{k_-}(t) - v_{k_-}(t)
c_{k_-}^\dag c_{-k_-}^\dag}|{\rm vac}\rangle,
\label{psimm}
\end{align}
where the state $|\text{vac}\ra$  is annihilated by all $c_k$
operators and  time evolution of the Bogolubov modes is governed by
\be
\ii\frac{d}{dt}
\left(
\begin{array}{c}
v_k \\
u_k
\end{array}
\right)
=2\left(
\begin{array}{cc}
g-\cos(k) & -\sin(k) \\
-\sin(k)  & \cos(k)-g
\end{array}
\right)
\left(
\begin{array}{c}
v_{k} \\
u_{k}
\end{array}
\right).
\label{Bogt}
\ee

Computation of the time-dependent longitudinal magnetization in our periodic system is now reduced to evaluation 
of 
\be
M_x(t)=\la\psi_-(t)|c_1+c_1^\dag|\psi_+(t)\ra+\cc
\label{eded}
\ee
If we now put (\ref{psipp}) and  (\ref{psimm}) into (\ref{eded}) and then 
transform  $c_{k_\pm}$ operators to the position space inverting (\ref{Fou})--so that all
operators are defined  on the same Hilbert space--we will quickly realize how
complicated the resulting  expression is. This obstacle discourages us from
using the free fermion representation in our studies of periodic chains.

Finally, having (\ref{Bogt}) at hand, one can find by diagonalization of the $2\times2$
Hamiltonian that the energy gap for excitation of the pair of $\pm k$ fermionic modes is
\be
4\sqrt{g^2-2g\cos(k)+1}.
\ee
The smallest value of that gap, in each of the parity subspaces, is 
the dynamical gap $\Delta_\pm$--see also Fig. \ref{levels_fig}. Thereby,
\be
\Delta_+=4\sqrt{g^2-2g\cos(\pi/N)+1}, \ \Delta_-=4\sqrt{g^2-2g\cos(2\pi/N)+1}.
\label{DD}
\ee

\section{Diagonalization and time evolution of Ising chains with open boundaries}
\label{OBC_app}
We will briefly summarize here technicalities related to diagonalization
of the  Ising chain with open boundaries, computation of its longitudinal magnetization, and
its time evolution.

{\bf Diagonalization}. We follow here \cite{YoungPRB1997} providing  an early take on this subject. 
The diagonalization begins  with Jordan-Wigner spin-to-fermion mapping (\ref{JW})
transforming   Hamiltonian (\ref{OBC}) to the following quadratic form
\be
\begin{aligned}
&\tilde H = \Psi^{\dagger} {\cal H} \Psi,\\
&\Psi^{\dagger} = (c_1^{\dagger} \ldots c_N^{\dagger} \, c_1 \ldots c_N),\\
&{\cal H} = \left ( \begin{array}{cc}
A & B  \\
-B & -A \\
\end{array} \right),
\end{aligned}
\label{OBCnext}
\ee
where $A$ and $B$ are $N \times N$ tridiagonal matrices
\be
\begin{aligned}
&A_{ij} = g \, \delta_{i,j} - 1/2 \, \delta_{i,j+1} -1/2 \, \delta_{i+1,j},\\ 
&B_{ij} = 1/2 \, \delta_{i,j+1} -1/2 \, \delta_{i+1,j}.
\end{aligned}
\label{aaabbb}
\ee
We mention in passing that we have corrected a misprint from  \cite{YoungPRB1997} in the expression for
$B_{ij}$. 

Next,  for every value of the magnetic field $g$, one  can perform the Bogolubov transformation
\be
\begin{aligned}
&\Psi = \beta  \Gamma,\\
&\Gamma^\dag = (\gamma_1^{\dagger} \ldots \gamma_N^{\dagger} \, \gamma_1 \ldots \gamma_N)
\end{aligned}
\ee
choosing real orthogonal matrix $\beta$ in  such a way that (\ref{OBCnext}) is diagonal. This leads
to (\ref{Hbog}) and the related equations (\ref{dE1})--(\ref{pppp}). Two remarks are in order now.

First, as a self-consistency check of our calculations, we have 
verified  that energies of single-particle excitations, which  we have
obtained
from numerical 
diagonalization of ${\cal H}$, very well  agree  with the results coming from 
(\ref{Ei}) combined with (\ref{pppp}).

Second, the symmetry-breaking gap can be analytically  calculated 
from (\ref{pppp}) only at the critical point, where 
\begin{equation}
\delta(g=1)  = 4 \sin \frac{\pi}{4N+2} \approx \frac{\pi}{N}.
\label{gapp}
\end{equation}
This  is about twice larger than the symmetry-breaking gap in the periodic
chain  (\ref{gapcritical}).

{\bf Equilibrium longitudinal magnetization.} 
To compute the equilibrium longitudinal magnetization $M_x^\eq$, we evaluate 
(\ref{Mx_open}) in the state 
\be
\frac{\vert GS \rangle + \gamma_1^{\dagger} \vert GS \rangle}{\sqrt{2}},
\ee
where $\vert GS\rangle$ is the ground state annihilated by all $\gamma_i$ operators. 
This leads to the following expression for the longitudinal magnetization 
after employment of  Wick's theorem  
\begin{equation}
M_x^\eq = \langle GS \vert \gamma_1 \,  \sigma^x_{N/2} \vert GS \rangle.
\end{equation}
The operator $\gamma_1 \sigma^x_{N/2}$ can be conveniently written as
\begin{equation}
\gamma_1 \sigma^x_{N/2} = \gamma_1 \, a_{N/2} \, b_{N/2-1} a_{N/2-1} \ldots b_1 \, a_1, \ a_i = c_i + c_i^{\dagger},
\ b_i = c_i - c_i^{\dagger}. 
\end{equation}
Using Wick's theorem again, one can show that 
\be
\begin{aligned}
\label{eq:pfaffian}
&\langle GS \vert \gamma_1 \,  \sigma^x_{N/2} \vert GS \rangle = \text{Pf}(G),\\
&G=\left(
\begin{array}{cccccc}
	0 &  \two{\gamma_1}{a_{N/2}} & \two{\gamma_1}{b_{N/2-1}} & \two{\gamma_1}{a_{N/2-1}} & \cdots & \two{\gamma_1}{a_1}\\
	&  0          & \two{a_{N/2}}{b_{N/2-1}} & \two{a_{N/2}}{a_{N/2-1}} & \cdots & \two{a_{N/2}}{a_1}\\
	&             & 0          & \two{b_{N/2-1}}{a_{N/2-1}} & \cdots & \two{b_{N/2-1}}{a_1}\\
	&             &            &            &        &   \vdots\\
	&             &            &            &        &  0
	\end{array}
	\right),
\end{aligned}
\ee
where $\text{Pf}$  stands for Pfaffian, the lower triangle of the $G$ matrix can be obtained by the relation $G=-G^T$, and 
the expectation values  are calculated in the ground state $|GS\ra$. 
Pfaffians of  skew-symmetric matrices can be efficiently  computed using Hausholder transformation \cite{wimmer}. 

{\bf Time evolution}. We work in the Heisenberg picture. Our evolutions start
at  time $t_0$ from the equal superposition of the two  lowest-energy eigenstates of
$\tilde H[g(t_0)]$
\be
\frac{\vert GS[g(t_0)]\rangle + \gamma_1^{\dagger} \vert GS[g(t_0)] \rangle}{\sqrt{2}}.
\label{rrrr}
\ee
Two remarks are in order now.

First,  initial state (\ref{rrrr})  for time evolution is constructed in the same way as 
for  evolutions in periodic chains.
In particular, the two states in  (\ref{rrrr}) have different
parities. In fact, it is perhaps worth to say again  that  Hamiltonian $\tilde H$ 
for the  Ising chain with open boundaries  commutes with the parity operator.
Therefore, its
eigenstates can be labeled with the $\pm1$  parities.
 Moreover,  expectation values of the $\sigma^x_i$ operators in all
eigenstates of $\tilde H$ are zero. The very same properties are found  in  periodic
chains, which have been  discussed in Secs. \ref{Idea_sec}--\ref{Periodic_sec}. 

Second, using  quench protocol (\ref{godt}), one gets $g(t_0=-g_f\tau_Q)=0$ and the initial state (\ref{rrrr}) is 
given by (\ref{initial}).  It is numerically convenient 
for us, however,  to begin  evolutions from the  slightly non-zero $g$, which
we do by choosing $t_0$ such that $g(t_0)=0.001$. 

Time-dependent longitudinal magnetization  (\ref{Mx_open})  is
then  expressed as
\be
M_x(t) = \text{Re}\langle GS[g(t_0)] \vert \gamma_1 \,  \sigma^x_{N/2}(t) \vert GS[g(t_0)] \rangle,
\label{Mxt}
\ee
where the operator $\gamma_1$ is defined at time $t_0$.
The matrix element in this equation can be  computed just as (\ref{eq:pfaffian}) expect $a_i$ and $b_i$
operators are now time dependent. Thus,  we need to know their time evolution,
which can be extracted from 
\begin{equation}
\Psi(t) = U(t)\Psi(t_0),
\end{equation}
where the $N\times N$ unitary matrix $U(t)$ can be obtained by solving 
\be
\frac{d}{dt} U(t) = -2\ii {\cal H}  U(t)
\label{kkklll}
\ee
with the initial condition $U(t_0)=\mathbb{1}$. Equation (\ref{kkklll}) can be derived from 
the Heisenberg equations for the $c_i(t)$ and $c_i^\dag(t)$ operators.
We solve it numerically with the  Suzuki-Trotter method of order two with the 
time step smaller or equal to $0.01$ \cite{trotter}. 
We have checked that such a procedure produces well-converged results.

Having $U(t)$ and the Bogolubov matrix $\beta$ at the  time $t_0$,
we can relate $c_i(t)$ and $c_i^\dag(t)$ operators to $\gamma_i$ and $\gamma_i^\dag$ appearing 
in the diagonal form of $\tilde H[g(t_0)]$. 
Namely,
\begin{equation}
\Psi(t) = W(t)  \Gamma, \  W(t) = U(t)  \beta,
\label{aassww}
\end{equation}
where the  matrix $W$ has the following structure
\begin{equation}
W = 
\left(\begin{array}{cc}
C & D\\
\overline{D} & \overline{C}
\end{array}\right)
\end{equation}
with $C$ and $D$ being $N\times N$ complex matrices.
Transformation (\ref{aassww})  can be used  to compute all  correlation functions, 
from time-dependent version of (\ref{eq:pfaffian}), 
needed for getting  (\ref{Mxt}).
For example, after straightforward manipulations one can show that 
\begin{equation} 
\two{a_i}{a_j} = \sum_{k=1}^N \B{
C_{ik}  D_{jk} + 
C_{ik}  \overline{C_{jk}} + 
\overline{D_{ik}}  D_{jk} +  
\overline{D_{ik}} \, \overline{C_{jk}}\,}.
\end{equation}


\begin{thebibliography}{48}
\expandafter\ifx\csname natexlab\endcsname\relax\def\natexlab#1{#1}\fi
\expandafter\ifx\csname bibnamefont\endcsname\relax
  \def\bibnamefont#1{#1}\fi
\expandafter\ifx\csname bibfnamefont\endcsname\relax
  \def\bibfnamefont#1{#1}\fi
\expandafter\ifx\csname citenamefont\endcsname\relax
  \def\citenamefont#1{#1}\fi
\expandafter\ifx\csname url\endcsname\relax
  \def\url#1{\texttt{#1}}\fi
\expandafter\ifx\csname urlprefix\endcsname\relax\def\urlprefix{URL }\fi
\providecommand{\bibinfo}[2]{#2}
\providecommand{\eprint}[2][]{\url{#2}}

\bibitem[{Sac({\natexlab{a}})}]{Sachdev}
\bibinfo{note}{S. Sachdev {\it {Q}uantum {P}hase {T}ransitions} (Cambridge
  University Press, 2011).}

\bibitem[{Con()}]{ContinentinoBook}
\bibinfo{note}{M. Continentino {\it Quantum Scaling in Many-Body Systems: An
  Approach to Quantum Phase Transitions} (Cambridge University Press, 2nd
  edition, 2017).}

\bibitem[{Sac({\natexlab{b}})}]{SachdevToday}
\bibinfo{note}{S. Sachdev and B. Keimer, Phys. Today {\bf 64}, 29 (2011).}

\bibitem[{Lew()}]{LewensteinAdv}
\bibinfo{note}{M. Lewenstein, A. Sanpera, V. Ahufinger, B. Damski, A. Sen De,
  and U. Sen, Adv. Phys. {\bf 56}, 243 (2007).}

\bibitem[{Jac({\natexlab{a}})}]{JacekPRL2005}
\bibinfo{note}{J. Dziarmaga, Phys. Rev. Lett. {\bf 95}, 245701 (2005).}

\bibitem[{Dor()}]{DornerPRL2005}
\bibinfo{note}{W. H. Zurek, U. Dorner, and P. Zoller, Phys. Rev. Lett. {\bf
  95}, 105701 (2005).}

\bibitem[{Pol({\natexlab{a}})}]{PolkovnikovPRB2005}
\bibinfo{note}{A. Polkovnikov, Phys. Rev. B {\bf 72}, 161201(R) (2005).}

\bibitem[{Ral()}]{RalfPRA2007}
\bibinfo{note}{S. Mostame, G. Schaller, and R. Sch\"utzhold, Phys. Rev. A {\bf
  76}, 030304(R) (2007).}

\bibitem[{Pol({\natexlab{b}})}]{PolkovnikovPRL2008}
\bibinfo{note}{R. Barankov and A. Polkovnikov, Phys. Rev. Lett. {\bf 101},
  076801 (2008).}

\bibitem[{Sen({\natexlab{a}})}]{SenPRB2009}
\bibinfo{note}{S. Mondal, K. Sengupta, and D. Sen, Phys. Rev. B {\bf 79},
  045128 (2009).}

\bibitem[{San({\natexlab{a}})}]{SantoroPRB2009}
\bibinfo{note}{D. Patan\`e, L. Amico, A. Silva, R. Fazio, and G. E. Santoro,
  Phys. Rev. B {\bf 80}, 024302 (2009).}

\bibitem[{Jac({\natexlab{b}})}]{JacekPRA2007}
\bibinfo{note}{L. Cincio, J. Dziarmaga, M. M. Rams, and W. H. Zurek, Phys. Rev.
  A {\bf 75}, 052321 (2007).}

\bibitem[{Sen({\natexlab{b}})}]{SenPRA2009}
\bibinfo{note}{K. Sengupta and D. Sen, Phys. Rev. A {\bf 80}, 032304 (2009).}

\bibitem[{Arn()}]{ArnabPRB2010}
\bibinfo{note}{A. Das, Phys. Rev. B {\bf 82}, 172402 (2010).}

\bibitem[{Kol()}]{KolodrubetzPRL2012}
\bibinfo{note}{M. Kolodrubetz, B. K. Clark, and D. A. Huse, Phys. Rev. Lett.
  {\bf 109}, 015701 (2012).}

\bibitem[{Fra()}]{FrancuzPRB2016}
\bibinfo{note}{A. Francuz, J. Dziarmaga, B. Gardas, and W. H. Zurek, Phys. Rev.
  B {\bf 93}, 075134 (2016).}

\bibitem[{San({\natexlab{b}})}]{SantoroJstat2015}
\bibinfo{note}{A. Russomanno, S. Sharma, A. Dutta, and G. E. Santoro, J. Stat.
  Mech. (2015) P08030.}

\bibitem[{Pus()}]{PuskarovSciP2016}
\bibinfo{note}{T. Puskarov and D. Schuricht, SciPost Phys. {\bf 1}, 003
  (2016).}

\bibitem[{Apo()}]{ApollaroSciRep2017}
\bibinfo{note}{S. Lorenzo, J. Marino, F. Plastina, G. M. Palma, and T. J. G.
  Apollaro, Sci. Rep. {\bf 7}, 5672 (2017).}

\bibitem[{Mic()}]{MichalOne}
\bibinfo{note}{M. Bia{\l}o{\'{n}}czyk and B. Damski, J. Stat. Mech. (2018)
  073105.}

\bibitem[{Ado()}]{AdolfoPRL2018}
\bibinfo{note}{A. del Campo, Phys. Rev. Lett. {\bf 121}, 200601 (2018).}

\bibitem[{Mar()}]{MarekArxiv}
\bibinfo{note}{M. M. Rams, J. {Dziarmaga}, and W. H. {Zurek}, Phys. Rev. Lett.
  {\bf 123}, 130603 (2019).}

\bibitem[{Blo()}]{BlochSci2017}
\bibinfo{note}{C. Gross and I. Bloch, Science {\bf 357}, 995 (2017).}

\bibitem[{LMG()}]{LMGexp}
\bibinfo{note}{V. Makhalov, T. Satoor, A. Evrard, T. Chalopin, R. Lopes, and S.
  Nascimbene, Phys. Rev. Lett. {\bf 123}, 120601 (2019).}

\bibitem[{Lie()}]{Lieb1961}
\bibinfo{note}{E. Lieb, T. Schultz, and D. Mattis, Ann. Phys. (N.Y.) {\bf 16},
  407 (1961).}

\bibitem[{Pfe()}]{Pfeuty}
\bibinfo{note}{P. Pfeuty, Ann. Phys. {\bf 57}, 79 (1970).}

\bibitem[{Wic()}]{Wick}
\bibinfo{note}{G. F. Giuliani and G. Vignale, {\it Quantum Theory of the
  Electron Liquid} (Cambridge University Press, 2005).}

\bibitem[{BDJ()}]{BDJPA2014}
\bibinfo{note}{B. Damski and M. M. Rams, J. Phys. A {\bf 47}, 025303 (2014).}

\bibitem[{BDP()}]{BDPRL2005}
\bibinfo{note}{B. Damski, Phys. Rev. Lett. {\bf 95}, 035701 (2005).}

\bibitem[{Kib()}]{KibbleRev}
\bibinfo{note}{T. W. B. Kibble, Phys. Rep. {\bf 67},183 (1980).}

\bibitem[{Zur()}]{ZurekRev}
\bibinfo{note}{W. H. Zurek, Phys. Rep. {\bf 276}, 177 (1996).}

\bibitem[{del({\natexlab{a}})}]{delCampoReva}
\bibinfo{note}{A. del Campo, T. W. B. Kibble, and W. H. Zurek, J. Phys.:
  Condens. Matter {\bf 25}, 404210 (2013).}

\bibitem[{del({\natexlab{b}})}]{delCampoRevb}
\bibinfo{note}{A. del Campo and W. H. Zurek, Int. J. Mod. Phys. A {\bf 29},
  1430018 (2014).}

\bibitem[{Jac({\natexlab{c}})}]{JacekAdv2010}
\bibinfo{note}{J. Dziarmaga, Adv. Phys. {\bf 59}, 1063 (2010).}

\bibitem[{Mat()}]{Mathematica}
\bibinfo{note}{Wolfram Research, Inc., Mathematica, Version 11.0, Champaign, IL
  (2016).}

\bibitem[{Kor()}]{KorenblitNJP2012}
\bibinfo{note}{S. Korenblit {\it et al.}, New J. Phys. {\bf 14} 095024 (2012).}

\bibitem[{Cab()}]{CabreraPRB1987}
\bibinfo{note}{G. G. Cabrera and R. Jullien, Phys. Rev. B {\bf 35}, 7062
  (1987).}

\bibitem[{Osk()}]{Oskar}
\bibinfo{note}{O. A. Pro\'sniak, Phys. Scr. {\bf 94}, 085201 (2019).}

\bibitem[{Luk({\natexlab{a}})}]{LukinNature2017}
\bibinfo{note}{H. Bernien {\it et al.}, Nature {\bf 551}, 579 (2017).}

\bibitem[{Mon()}]{MonroeNature2017}
\bibinfo{note}{J. Zhang, G. Pagano, P. W. Hess, A. Kyprianidis, P. Becker, H.
  Kaplan, A. V. Gorshkov, Z.-X. Gong, and C. Monroe, Nature {\bf 551}, 601
  (2017).}

\bibitem[{Luk({\natexlab{b}})}]{LukinNature2019}
\bibinfo{note}{A. Keesling {\it et al.}, Nature {\bf 568}, 207 (2019).}

\bibitem[{Por()}]{PorrasPRL2004}
\bibinfo{note}{D. Porras and J. I. Cirac, Phys. Rev. Lett. {\bf 92}, 207901
  (2004).}

\bibitem[{Sch()}]{SchaussQST2018}
\bibinfo{note}{P. Schauss, Quantum Sci. Technol. {\bf 3}, 023001 (2018).}

\bibitem[{Har()}]{HarrisSci2018}
\bibinfo{note}{R. Harris {\it et al.}, Science {\bf 361}, 162 (2018).}

\bibitem[{Jac({\natexlab{d}})}]{JacekSciRep2018}
\bibinfo{note}{B. Gardas, J. Dziarmaga, W. H. Zurek, and M. Zwolak, Sci. Rep.
  {\bf 8}, 4539 (2018).}

\bibitem[{You()}]{YoungPRB1997}
\bibinfo{note}{A. P. Young, Phys. Rev. B {\bf 56}, 11691 (1997).}

\bibitem[{wim()}]{wimmer}
\bibinfo{note}{M. Wimmer, ACM Trans. Math. Software {\bf 38}, 30 (2012).}

\bibitem[{tro()}]{trotter}
\bibinfo{note}{N. Hatano and M. Suzuki, Finding Exponential Product Formulas of
  Higher Orders, in {\it Quantum Annealing and Other Optimization Methods},
  edited by A. Das and B. K. Chakrabarti (Springer, 2005), pp. 37-68.}

\end{thebibliography}

\end{document}